\documentclass[longauth]{aa} %

\usepackage{graphicx}
\usepackage{txfonts}
\usepackage[breaklinks=true]{hyperref}
\makeatletter
\renewcommand*\aa@pageof{, page \thepage{} of \pageref*{LastPage}}
\usepackage{amsmath}    %
 \usepackage{amssymb}   %
\usepackage{natbib}
\usepackage{colortbl}
\graphicspath{{./}{Figures/}}

\newcommand{\HI}{\textrm{H~{\textsc{i}}}}
\newcommand{\HII}{\textrm{H~{\textsc{ii}}}}
\newcommand{\jwst}{\textit{JWST}}
\newcommand{\mum}{$\mu$m\xspace}
\newcommand{\molh}{H$_2$\xspace}

\usepackage{color}

\definecolor{lightgrey}{rgb}{0.84, 0.84, 0.84}

\definecolor{cbpurple}{rgb}{0.47, 0.37, 0.94}

\definecolor{teal}{rgb}{0, .5, .5}

\definecolor{ballblue}{rgb}{0, 0, 0.8}

\definecolor{purple}{rgb}{.8, 0, 1}

\definecolor{bostonuniversityred}{rgb}{0.8, 0.0, 0.0}

\definecolor{capri}{rgb}{0.0, 0.75, 1.0}

\usepackage{soul}
\definecolor{darkorange}{rgb}{1.0, 0.55, 0.0}

\definecolor{olive}{rgb}{0.5, 0.5, 0}

\makeatletter
\renewcommand*\maketitle{%
  \thispagestyle{firstpage}
\begingroup
    \if@wideboxfn
    \setlength\bibindent{1.4\parindent}
    \else
    \setlength\bibindent{\parindent}
    \fi
    \renewcommand*\thefootnote{\@fnsymbol\c@footnote}%
    \renewcommand\@makefntext[1]{%
    \ifaa@longfn\hsize\textwidth\fi
    \noindent
    \hb@xt@\bibindent{\hss\@makefnmark\enspace}##1}
  \ifaa@twocolumn
  \begin{aa@strip}
    \aa@maketitle
  \end{aa@strip}
  \@thanks %
  \else
    \begingroup
      \let\thanks\footnote
      \aa@maketitle
    \endgroup
  \fi
\endgroup
  \setcounter{footnote}{0}%
}
\makeatother

\begin{document} 

\title{PDRs4All
\\  IV. An embarrassment of riches: \\ Aromatic infrared bands in the Orion Bar }
\titlerunning{PDRs4All IV: the Aromatic Infrared Bands in the Orion Bar}

\author{Ryan Chown \inst{1,2} \and
Ameek Sidhu \inst{1,2} \and
Els Peeters \inst{1,2,3} \and
Alexander G.~G.~M. Tielens \inst{4,5} \and
Jan Cami \inst{1,2,3} \and
Olivier Bern\'{e} \inst{6} \and
{E}milie Habart\inst{7} \and
Felipe Alarc\'on \inst{8} \and
Am\'elie Canin \inst{6} \and
Ilane Schroetter \inst{6} \and
Boris Trahin \inst{7} \and
Dries Van De Putte \inst{9} \and
Alain Abergel \inst{7}  \and
Edwin A. Bergin \inst{8} \and
Jeronimo Bernard-Salas \inst{10,11} \and
Christiaan Boersma \inst{12} \and
Emeric Bron \inst{13} \and
Sara Cuadrado \inst{14} \and
Emmanuel Dartois \inst{15} \and
Daniel Dicken \inst{16} \and
Meriem El-Yajouri \inst{7} \and
Asunci\'on Fuente \inst{17} \and
Javier R. Goicoechea \inst{14} \and
Karl D.\ Gordon \inst{9,18} \and
Lina Issa \inst{6} \and 
Christine Joblin \inst{4} \and
Olga Kannavou \inst{7} \and
Baria Khan \inst{1} \and
Ozan Lacinbala \inst{19} \and
David Languignon \inst{13} \and
Romane Le Gal \inst{20,21} \and
Alexandros Maragkoudakis \inst{12} \and
Raphael Meshaka \inst{7,13} \and
Yoko Okada \inst{22} \and
Takashi Onaka \inst{23, 24} \and
Sofia Pasquini \inst{1} \and
Marc W. Pound \inst{5} \and
Massimo Robberto \inst{9, 25} \and
Markus R\"ollig \inst{26,27} \and
Bethany Schefter \inst{1,2} \and
Thi\'{e}baut Schirmer \inst{7, 28} \and
S\'ilvia Vicente \inst{29} \and
Mark G. Wolfire \inst{5} \and 
Marion Zannese \inst{7} \and
Isabel Aleman \inst{30} \and
Louis Allamandola \inst{12, 31} \and
Rebecca Auchettl \inst{32} \and
Giuseppe Antonio Baratta \inst{33} \and
Salma Bejaoui \inst{12} \and
Partha P. Bera \inst{12,29} \and
John~H.~Black \inst{28} \and
Francois~Boulanger \inst{34} \and
Jordy Bouwman \inst{35, 36, 37} \and
Bernhard Brandl \inst{4,38} \and
Philippe Brechignac \inst{15} \and
Sandra Br\"unken \inst{39} \and
Mridusmita Buragohain \inst{40} \and
Andrew Burkhardt \inst{41} \and
Alessandra Candian \inst{42} \and
St\'{e}phanie Cazaux \inst{43} \and
Jose Cernicharo \inst{14} \and
Marin Chabot \inst{44} \and
Shubhadip Chakraborty \inst{45, 46} \and
Jason Champion \inst{6} \and
Sean W.J. Colgan \inst{12} \and
Ilsa R. Cooke \inst{47} \and
Audrey Coutens \inst{6} \and
Nick L.J. Cox \inst{10,11} \and
Karine Demyk \inst{6} \and
Jennifer Donovan Meyer \inst{48} \and
Sacha Foschino \inst{6} \and
Pedro Garc\'ia-Lario \inst{49} \and
Lisseth Gavilan \inst{12} \and
Maryvonne Gerin \inst{50} \and
Carl A. Gottlieb \inst{51} \and
Pierre Guillard \inst{52,53} \and
Antoine Gusdorf \inst{34,50} \and
Patrick Hartigan \inst{54} \and
Jinhua He \inst{55,56,57} \and
Eric Herbst \inst{58} \and
Liv Hornekaer \inst{59} \and
Cornelia J\"ager \inst{60} \and
Eduardo Janot-Pacheco\inst{61} \and
Michael Kaufman\inst{62} \and
Francisca Kemper\inst{63, 64, 65} \and
Sarah Kendrew\inst{66} \and
Maria S. Kirsanova\inst{67} \and
Pamela Klaassen\inst{16} \and
Sun Kwok\inst{68} \and
\'Alvaro Labiano \inst{69} \and
Thomas S.-Y. Lai \inst{70} \and
Timothy J. Lee \inst{12} \thanks{Tim Lee sadly passed away on Nov 3, 2022.} \and
Bertrand Lefloch \inst{20} \and
Franck Le Petit \inst{13} \and
Aigen Li \inst{71} \and
Hendrik Linz \inst{72} \and
Cameron J. Mackie \inst{73,74} \and
Suzanne C. Madden \inst{75} \and
Jo\"elle Mascetti \inst{76} \and
Brett A. McGuire \inst{48, 77} \and
Pablo Merino \inst{78} \and
Elisabetta R. Micelotta \inst{79} \and
Karl Misselt \inst{80} \and
Jon A. Morse \inst{81} \and
Giacomo Mulas \inst{33,6} \and
Naslim Neelamkodan \inst{82} \and
Ryou Ohsawa \inst{83} \and
Alain Omont \inst{84} \and
Roberta Paladini \inst{85} \and
Maria Elisabetta Palumbo \inst{33} \and
Amit Pathak \inst{86} \and
Yvonne J. Pendleton \inst{87} \and
Annemieke Petrignani \inst{88} \and
Thomas Pino \inst{15} \and
Elena Puga \inst{66} \and
Naseem Rangwala \inst{12} \and
Mathias Rapacioli \inst{89} \and
Alessandra Ricca \inst{12,3} \and
Julia Roman-Duval \inst{9} \and
Joseph Roser \inst{3,12} \and
Evelyne Roueff \inst{13} \and
Ga\"el Rouill\'e \inst{90} \and
Farid Salama \inst{12} \and
Dinalva A. Sales \inst{91} \and
Karin Sandstrom \inst{92} \and
Peter Sarre \inst{93} \and
Ella Sciamma-O'Brien \inst{12} \and
Kris Sellgren \inst{94} \and
Sachindev S. Shenoy \inst{95} \and
David Teyssier \inst{69} \and
Richard D. Thomas \inst{96} \and
Aditya Togi \inst{72} \and
Laurent Verstraete \inst{7} \and
Adolf N. Witt \inst{97} \and
Alwyn Wootten \inst{48} \and
Henning Zettergren \inst{96} \and
Yong Zhang \inst{98} \and
Ziwei E. Zhang \inst{99} \and
Junfeng Zhen \inst{100}
}

\institute{Department of Physics \& Astronomy, The University of Western Ontario, London ON N6A 3K7, Canada; \email{rchown3@uwo.ca} \and
Institute for Earth and Space Exploration, The University of Western Ontario, London ON N6A 3K7, Canada    \and
Carl Sagan Center, SETI Institute, 339 Bernardo Avenue, Suite 200, Mountain View, CA 94043, USA \and
Leiden Observatory, Leiden University, P.O. Box 9513, 2300 RA Leiden, The Netherlands \and
Astronomy Department, University of Maryland, College Park, MD 20742, USA \and
Institut de Recherche en Astrophysique et Plan\'etologie, Universit\'e Toulouse III - Paul Sabatier, CNRS, CNES, 9 Av. du colonel Roche, 31028 Toulouse Cedex 04, France \and
Institut d'Astrophysique Spatiale, Universit\'e Paris-Saclay, CNRS,  B$\hat{a}$timent 121, 91405 Orsay Cedex, France \and
Department of Astronomy, University of Michigan, 1085 South University Avenue, Ann Arbor, MI 48109, USA \and
Space Telescope Science Institute, 3700 San Martin Drive, Baltimore, MD, 21218, USA \and
ACRI-ST, Centre d’Etudes et de Recherche de Grasse (CERGA), 10 Av. Nicolas Copernic, F-06130 Grasse, France \and
INCLASS Common Laboratory., 10 Av. Nicolas Copernic, 06130 Grasse, France \and
NASA Ames Research Center, MS 245-6, Moffett Field, CA 94035-1000, USA \and
LERMA, Observatoire de Paris, PSL Research University, CNRS, Sorbonne Universit\'es, F-92190 Meudon, France \and
Instituto de F\'{\i}sica Fundamental  (CSIC),  Calle Serrano 121-123, 28006, Madrid, Spain \and
Institut des Sciences Mol\'eculaires d'Orsay, CNRS, Universit\'e Paris-Saclay,  B$\hat{a}$timent 520, 91405 Orsay Cedex, France \and
UK Astronomy Technology Centre, Royal Observatory Edinburgh, Blackford Hill, Edinburgh EH9 3HJ, UK \and
Observatorio Astron\'{o}mico Nacional (OAN,IGN), Alfonso XII, 3, E-28014 Madrid, Spain \and
Sterrenkundig Observatorium, Universiteit Gent, Gent, Belgium \and
KU Leuven Quantum Solid State Physics (QSP), Celestijnenlaan 200d - box 2414, 3001 Leuven, Belgium \and
Institut de Plan\'etologie et d'Astrophysique de Grenoble (IPAG), Universit\'e Grenoble Alpes, CNRS, F-38000 Grenoble, France \and
Institut de Radioastronomie Millim\'etrique (IRAM), 300 Rue de la Piscine, F-38406 Saint-Martin d'H\`{e}res, France \and
I. Physikalisches Institut der Universit\"{a}t zu K\"{o}ln, Z\"{u}lpicher Stra{\ss}e 77, 50937 K\"{o}ln, Germany \and
Department of Astronomy, Graduate School of Science, The University of Tokyo, 7-3-1 Bunkyo-ku, Tokyo 113-0033, Japan \and
Department of Physics, Faculty of Science and Engineering, Meisei University, 2-1-1 Hodokubo, Hino, Tokyo 191-8506, Japan \and
Johns Hopkins University, 3400 N. Charles Street, Baltimore, MD, 21218, USA \and
Physikalischer Verein - Gesellschaft für Bildung und Wissenschaft, Robert-Mayer-Straße 2, 60325 Frankfurt am Main, Germany \and
Institut für Angewandte Physik. Goethe-Universit{\"a}t Frankfurt, Max-von-Laue-Str. 1, 60438 Frankfurt am Main, Germany \and
Department of Space, Earth and Environment, Chalmers University of Technology, Onsala Space Observatory, SE-439 92 Onsala, Sweden \and
Instituto de Astrof\'isica e Ci\^{e}ncias do Espa\c co, Tapada da Ajuda, Edif\'icio Leste, 2\,$^{\circ}$ Piso, P-1349-018 Lisboa, Portugal  \and
Instituto de Física e Química, Universidade Federal de Itajubá, Av. BPS 1303, Pinheirinho, 37500-903, Itajubá, MG, Brazil   \and
Bay Area Environmental Research Institute, Moffett Field, CA 94035, USA   \and
Australian Synchrotron, Australian Nuclear Science and Technology Organisation (ANSTO), Victoria, Australia   \and
INAF - Osservatorio Astrofisico di Catania, Via Santa Sofia 78, 95123 Catania, Italy    \and
Laboratoire de Physique de l'\'Ecole Normale Sup\'erieure, ENS, Universit\'e PSL, CNRS, Sorbonne Universit\'e, Universit\'e de Paris, 75005, Paris, France \and
Laboratory for Atmospheric and Space Physics, University of Colorado, Boulder, CO 80303, USA \and
Department of Chemistry, University of Colorado, Boulder, CO 80309, USA \and 
Institute for Modeling Plasma, Atmospheres, and Cosmic Dust (IMPACT), University of Colorado, Boulder, CO 80303, USA \and
Faculty of Aerospace Engineering, Delft University of Technology, Kluyverweg 1, 2629 HS Delft, The Netherlands \and 
FELIX Laboratory, Institute for Molecules and Materials, Radboud University, Toernooiveld 7, 6525ED Nijmegen, The Netherlands \and
School of Physics, University of Hyderabad, Hyderabad, Telangana 500046, India \and
Department of Earth, Environment, and Physics Worcester State University, Worcester, MA, USA \and
Anton Pannekoek Institute for Astronomy (API), University of Amsterdam, Science Park 904, 1098 XH Amsterdam, The Netherlands \and
Delft University of Technology, Delft, The Netherlands \and
Laboratoire de Physique des deux infinis Ir\`{e}ne Joliot-Curie, Universit\'e Paris-Saclay, CNRS/IN2P3,  B$\hat{a}$timent 104, 91405 Orsay Cedex, France \and
Department of Chemistry, GITAM school of Science, GITAM Deemed to be University, Bangalore, India \and
Institut de Physique de Rennes, UMR CNRS 6251, Universit{\'e} de Rennes 1, Campus de Beaulieu, 35042 Rennes Cedex, France \and
Department of Chemistry, The University of British Columbia, Vancouver, British Columbia, Canada \and
National Radio Astronomy Observatory (NRAO), 520 Edgemont Road, Charlottesville, VA 22903, USA \and
European Space Astronomy Centre (ESAC/ESA), Villanueva de la Ca\~nada, E-28692 Madrid, Spain \and
Observatoire de Paris, PSL University, Sorbonne Universit\'e, LERMA, 75014, Paris, France \and
Harvard-Smithsonian Center for Astrophysics, 60 Garden Street, Cambridge MA 02138, USA \and
Sorbonne Universit\'{e}, CNRS, UMR 7095, Institut d'Astrophysique de Paris, 98bis bd Arago, 75014 Paris, France   \and
Institut Universitaire de France, Minist\`{e}re de l'Enseignement Sup{\'e}rieur et de la Recherche, 1 rue Descartes, 75231 Paris Cedex 05, France \and
Department of Physics and Astronomy, Rice University, Houston TX, 77005-1892, USA \and
Yunnan Observatories, Chinese Academy of Sciences, 396 Yangfangwang, Guandu District, Kunming, 650216, China \and
Chinese Academy of Sciences South America Center for Astronomy, National Astronomical Observatories, CAS, Beijing 100101, China \and
Departamento de Astronom\'{i}a, Universidad de Chile, Casilla 36-D, Santiago, Chile \and
Departments of Chemistry and Astronomy, University of Virginia, Charlottesville, Virginia 22904, USA \and
InterCat and Dept. Physics and Astron., Aarhus University, Ny Munkegade 120, 8000 Aarhus C, Denmark \and
Laboratory Astrophysics Group of the Max Planck Institute for Astronomy at the Friedrich Schiller University Jena, Institute of Solid State Physics, Helmholtzweg 3, 07743 Jena, Germany   \and
Instituto de Astronomia, Geof\'isica e Ci\^encias Atmosf\'ericas, Universidade de S\~ao Paulo, 05509-090 S\~ao Paulo, SP, Brazil   \and
Department of Physics and Astronomy, San Jos\'e State University, San Jose, CA 95192, USA   \and
Institut de Ciencies de l’Espai (ICE, CSIC), Can Magrans, s/n, E-08193 Bellaterra, Barcelona, Spain    \and
ICREA, Pg. Llu{\`i}s Companys 23, E-08010 Barcelona, Spain   \and
Institut d’Estudis Espacials de Catalunya (IEEC), E-08034 Barcelona, Spain   \and
European Space Agency, Space Telescope Science Institute, 3700 San Martin Drive, Baltimore MD 21218, USA   \and
Institute of Astronomy, Russian Academy of Sciences, 119017, Pyatnitskaya str., 48 , Moscow, Russia   \and
Department of Earth, Ocean, \& Atmospheric Sciences, University of British Columbia, Canada V6T 1Z4   \and
Telespazio UK for ESA, ESAC, E-28692 Villanueva de la Ca\~nada, Madrid, Spain   \and
IPAC, California Institute of Technology, Pasadena, CA, USA   \and
Department of Physics and Astronomy, University of Missouri, Columbia, MO 65211, USA   \and
Max Planck Institute for Astronomy, K\"onigstuhl 17, 69117 Heidelberg, Germany   \and
Chemical Sciences Division, Lawrence Berkeley National Laboratory, Berkeley, California, USA   \and
Kenneth S.~Pitzer Center for Theoretical Chemistry, Department of Chemistry, University of California -- Berkeley, Berkeley, California, USA   \and
AIM, CEA, CNRS, Universit\'e Paris-Saclay, Universit\'e Paris Diderot, Sorbonne Paris Cit\'e, 91191 Gif-sur-Yvette, France   \and
Institut des Sciences Moléculaires, CNRS, Université de Bordeaux, 33405 Talence, France   \and
Department of Chemistry, Massachusetts Institute of Technology, Cambridge, MA 02139, USA   \and
Instituto de Ciencia de Materiales de Madrid (CSIC), Sor Juana Ines de la Cruz 3, E28049, Madrid, Spain   \and
Department of Physics, PO Box 64, 00014 University of Helsinki, Finland   \and
Steward Observatory, University of Arizona, Tucson, AZ 85721-0065, USA   \and
BoldlyGo Institute, 31 W 34TH ST FL 7 STE 7159, New York, NY 10001   \and
Department of Physics, College of Science, United Arab Emirates University (UAEU), Al-Ain, 15551, UAE   \and
National Astronomical Observatory of Japan, National Institutes of Natural Science, 2-21-1 Osawa, Mitaka, Tokyo 181-8588, Japan   \and
Sorbonne Universit\'{e}, CNRS, UMR 7095, Institut d'Astrophysique de Paris, 98bis bd Arago, 75014 Paris, France   \and
California Institute of Technology, IPAC, 770, S. Wilson Ave., Pasadena, CA 91125, USA   \and
Department of Physics, Institute of Science, Banaras Hindu University, Varanasi 221005, India   \and
University of Central Florida, Orlando, FL 32765 \and
Van’t Hoff Institute for Molecular Sciences, University of Amsterdam, PO Box 94157, 1090 GD, Amsterdam, The Netherlands    \and
Laboratoire de Chimie et Physique Quantiques LCPQ/FERMI, UMR5626, Universit\'e de Toulouse (UPS) and CNRS, Toulouse, France   \and
Laboratory Astrophysics Group of the Max Planck Institute for Astronomy at the Friedrich Schiller University Jena, Institute of Solid State Physics, Helmholtzweg 3, 07743 Jena, Germany   \and
Instituto de Matem\'atica, Estat\'istica e F\'isica, Universidade Federal do Rio Grande, 96201-900, Rio Grande, RS, Brazil   \and
Center for Astrophysics and Space Sciences, Department of Physics, University of California, San Diego, 9500 Gilman Drive, La Jolla, CA 92093, USA   \and
School of Chemistry, The University of Nottingham, University Park, Nottingham, NG7 2RD, United Kingdom   \and
Astronomy Department, Ohio State University, Columbus, OH 43210 USA   \and
Space Science Institute, 4765 Walnut St., R203, Boulder, CO 80301  \and
Department of Physics, Stockholm University, SE-10691 Stockholm, Sweden \and  
Department of Physics, Texas State University, San Marcos, TX 78666 USA   \and
School of Physics and Astronomy, Sun Yat-sen University, 2 Da Xue Road, Tangjia, Zhuhai 519000,  Guangdong Province, China   \and
Star and Planet Formation Laboratory, RIKEN Cluster for Pioneering Research, Hirosawa 2-1, Wako, Saitama 351-0198, Japan \and
Institute of Deep Space Sciences, Deep Space Exploration Laboratory, Hefei 230026, China }

   \date{Received 15 April 2023; accepted 1 August 2023}

  \abstract
  {Mid-infrared observations of photodissociation regions (PDRs) are dominated by strong emission features called aromatic infrared bands (AIBs). The most prominent AIBs are found at 3.3, 6.2, 7.7, 8.6, and 11.2~$\mu$m. The most sensitive, highest-resolution infrared spectral imaging data ever taken of the prototypical PDR, the Orion Bar, have been captured by  \jwst. These high-quality data allow for an unprecedentedly detailed view of AIBs. }
   {We provide an inventory of the AIBs found in the Orion Bar, along with mid-IR template spectra from five distinct regions in the Bar: the molecular PDR (i.e. the three \molh\ dissociation fronts), the atomic PDR, and the \HII\ region.}
   {We used \jwst\ NIRSpec IFU and MIRI MRS observations of the Orion Bar from the \jwst\ Early Release Science Program, PDRs4All (ID: 1288). We extracted five template spectra to represent the morphology and environment of the Orion Bar PDR. We investigated and characterised the AIBs in these template spectra. We describe the variations among them here.}
   {The superb sensitivity and the spectral and spatial resolution of these \jwst\ observations reveal many details of the AIB emission and enable an improved characterization of their detailed profile shapes and sub-components. The Orion Bar spectra are dominated by the well-known AIBs at 3.3, 6.2, 7.7, 8.6, 11.2, and 12.7~\mum\ with well-defined profiles. In addition, the spectra display a wealth of weaker features and sub-components. The widths of many AIBs show clear and systematic variations, being narrowest in the atomic PDR template, but showing a clear broadening in the \HII\ region template while the broadest bands are found in the three dissociation front templates. In addition, the relative strengths of AIB (sub-)components vary among the template spectra as well. All AIB profiles are characteristic of class A sources as designated by \citet{peeters2002}, except for the 11.2~\mum AIB profile deep in the molecular 
   zone, which belongs to class B$_{11.2}$. Furthermore, the observations show that the sub-components that contribute to the 5.75, 7.7, and 11.2~$\mu$m AIBs become much weaker in the PDR surface layers. We attribute this to the presence of small, more labile carriers in the deeper PDR layers that are photolysed away in the harsh radiation field near the surface. The 3.3/11.2 AIB intensity ratio decreases by about 40\% between the dissociation fronts and the \HII\ region, indicating a shift in the polycyclic aromatic hydrocarbon (PAH) size distribution to larger PAHs in the PDR surface layers, also likely  due to the effects of photochemistry. The observed broadening of the bands in the molecular PDR is consistent with an enhanced importance of smaller PAHs since smaller PAHs attain a higher internal excitation energy at a fixed photon energy. 
   }
   {Spectral-imaging observations of the Orion Bar using \jwst\ yield key insights into the photochemical evolution of PAHs, such as the evolution responsible for the shift of 11.2~\mum AIB emission from class B$_{11.2}$ in the molecular PDR to class A$_{11.2}$ in the PDR surface layers. This photochemical evolution is driven by the increased importance of FUV processing in the PDR surface layers, resulting in a ``weeding out'' of the weakest links of the PAH family in these layers. For now, these \jwst\ observations are consistent with a model in which the underlying PAH family is composed of a few species: the so-called `grandPAHs'.  }


   \maketitle
   
\section{Introduction}
\label{sec:intro}

A major component of the infrared (IR) emission near star-forming regions in the Universe consists of a set of broad emission features at 3.3, 6.2, 7.7, 8.6, 11.2, and 12.7~\mum\ \citep[e.g.][and references therein]{tielens:08}. These mid-IR emission features, referred to as aromatic infrared bands (AIBs), are generally attributed to vibrational emission from polycyclic aromatic hydrocarbons and related species upon absorption of interstellar far-ultraviolet (FUV; 6--13.6 eV) photons \citep{Leger:84, Allamandola:85}. The AIB spectrum is very rich and consists of the main bands listed above and a plethora of weaker emission features. Moreover, many AIBs are in fact blends of strong and weak bands \citep[e.g.][]{Peeters:colorado}. The AIB emission is known to vary from source to source and spatially within extended sources in terms of the profile and relative intensities of the features \citep[e.g.][]{joblin1996, hony2001, berne2007,  Sandstrom:10, Boersma:12, candian2012, Stock:17, peeters2017}. 
These remarkably widespread emission features have been described in many diverse astronomical sources, including protoplanetary disks \citep[e.g.][]{Meeus:01, vicente2013}, \HII\ regions \citep[e.g.][]{Bregman:hiivspn:89, Peeters:cataloog:02}, reflection nebulae \citep[e.g.][]{peeters2002, werner04}, planetary nebulae \citep[e.g.][]{Gillett:73, Bregman:hiivspn:89, Beintema:pahs:96}, the interstellar medium (ISM) of galaxies ranging from the Milky Way \citep{boulanger1996}, the Magellanic Clouds \citep[e.g.][]{Vermeij:02, Sandstrom:10}, starburst galaxies, luminous and ultra-luminous IR galaxies, and high-redshift galaxies \citep[e.g.][]{Genzel:98, Lutz:98, peeters2004, yan2005, galliano2008}, as well as in the harsh environments of galactic nuclei \citep[e.g.][]{Smith:07, esquej2014, jensen2017}.

A useful observational proxy for studying AIBs is the spectroscopic classification scheme devised by \citet{peeters2002} which classifies each individual AIBs based on their profile shapes and precise peak positions (classes A, B, and C). While the AIBs observed in a given source generally belong to the same class, this is not always the case. In particular, the classes in the 6 to 9~\mum region do not always correspond to those of the 3.3 and 11.2 AIBs \citep{vandiedenhoven2004}. 
Class A sources are the most common -- they exhibit the ``classical'' AIBs, with a 6.2~$\mu$m AIB that peaks between 6.19 and 6.23~$\mu$m, a 7.7~$\mu$m complex in which the 7.6~$\mu$m sub-peak is stronger than the 7.8~$\mu$m sub-peak, and the 8.6~$\mu$m feature peaks at 8.6~$\mu$m. Class B sources can be slightly redshifted compared to class A, while at the same time the 7.7~$\mu$m complex peaks between 7.8 and 8~$\mu$m. Class C sources show a very broad emission band peaking near 8.2~$\mu$m, and typically do not exhibit the 6.2 or 7.7~$\mu$m AIBs.

These three classes were found to show a strong correlation with the type of object considered. The most common AIB spectrum, class A, is identified in the spectra of photodissociation regions (PDRs), H{\sc ii} regions, reflection nebulae, the ISM, and galaxies. The most widely used template for class A sources has been the spectrum of the Orion Bar \citep{peeters2002, vandiedenhoven2004}. Class B sources are isolated Herbig Ae/Be stars and a few evolved stars; in fact, evolved star spectra can belong to either of the classes. Class C sources include post-AGB and Herbig Ae/Be stars, as well as a few T-Tauri disks \citep{peeters2002, bouwman2008, Shannon:19}. More recent work has developed analogous classification schemes for other AIBs and has included a new class D \citep[e.g.][]{vandiedenhoven2004, sloan2014, matsuura:14}.

Observed variations in AIBs reflect changes in the molecular properties of the species responsible for the AIB emission \citep[charge, size, and molecular structure; e.g.][]{joblin1996, berne2007,  Pilleri:12, Boersma:13, Candian:15, peeters2017,Robertson1986,Dartois2004,pino2008,Godard2011,Jones2013}, which are set by the local physical conditions \citep[including FUV radiation field strength, $G_{0}$, gas temperature, and density, $n(\mathrm{H})$; e.g.][]{Bakes:modelI:01, galliano2008, Pilleri:12, Pilleri:15, stock2016, schirmer2020, Schirmer2022, Sidhu:model:22, Knight:1333:22, murga2022}. The observed variability in  AIB emission thus implies that the population responsible for their emission is not static, but undergoes photochemical evolution. 

Observations using space-based IR observatories -- in particular the Short-Wavelength Spectrometer \citep[SWS;][]{sws} on board the \textit{Infrared Space Observatory} \citep[\textit{ISO};][]{Kessler:iso:96} and the Infrared Spectrograph \citep[IRS;][]{Houck:04} on board the \textit{Spitzer Space Telescope} \citep{werner04} -- have revealed the richness of AIBs \citep[for a review see e.g.][]{Peeters:colorado, tielens:08}. However, obtaining a full understanding of the photochemical evolution underlying AIBs has been limited by insufficient spatial and spectral resolution (e.g. \textit{Spitzer-IRS}) or by limited sensitivity and spatial resolution (e.g. \textit{ISO}/SWS) of these IR facilities.

\jwst\ is set to unravel the observed complexity of AIBs, as it offers access to the full wavelength range of importance for AIB studies at medium spectral resolution and at unprecedented spatial resolution and sensitivity. \jwst\ is able to resolve, for the first time, where and how the photochemical evolution of polycyclic aromatic related species, the carriers of AIBs, occurs while providing a detailed view of the resulting AIB spectral signatures. The PDRs4All \jwst\ Early Release Science Program observed the prototypical highly irradiated PDR, the Orion Bar \citep{pdrs4all, Habart:im, Peeters:nirspec}. The Orion Bar PDR has a G$_{0}$ which varies with position from about $1\times 10^{4}$ to $4\times 10^{4}$ Habings \citep[e.g.][]{Marconi:1998, Peeters:nirspec} and it has a gas density which varies from of a few $10^{4}$ cm$^{-3}$ in the atomic PDR to $\sim10^{6}$ cm$^{-3}$ in the molecular region \citep[e.g.][]{Parmar:1991, Tauber:1994, Young_Owl:2000, Bernard-Salas:2012, Goicoechea2016,Joblin:18, Habart:im}. Given the proximity of Orion \citep[414 pc;][]{menten2007}, the PDRs4All dataset takes full advantage of \jwst's spatial resolution to showcase the AIB emission in unprecedented detail. 

In this paper, we present five MIRI-MRS template spectra representing key regions of the Orion Bar PDR. Combined with corresponding \jwst\ NIRSpec-IFU template spectra \citep{Peeters:nirspec}, we present an updated inventory and characterization of the AIB emission in this important reference source. We describe the observations, data reduction, and the determination of the underlying continuum in our template spectra in Sect.~\ref{sec:data}. In Sect.~\ref{sec:results}, we give a detailed account of the observed AIB bands and sub-components along with their vibrational assignments. We compare our findings with previous works and discuss the AIB profiles and the AIB variability in the Orion Bar in Sect.~\ref{sec:discussion}. Finally, we summarize our results and narrate a picture of the origins and evolution of the AIB emission in Sect.~\ref{sec:conclusions}. 

\section{Data and data processing}
\label{sec:data}

\begin{figure*}
    \centering
    \resizebox{\hsize}{!}{
  \includegraphics{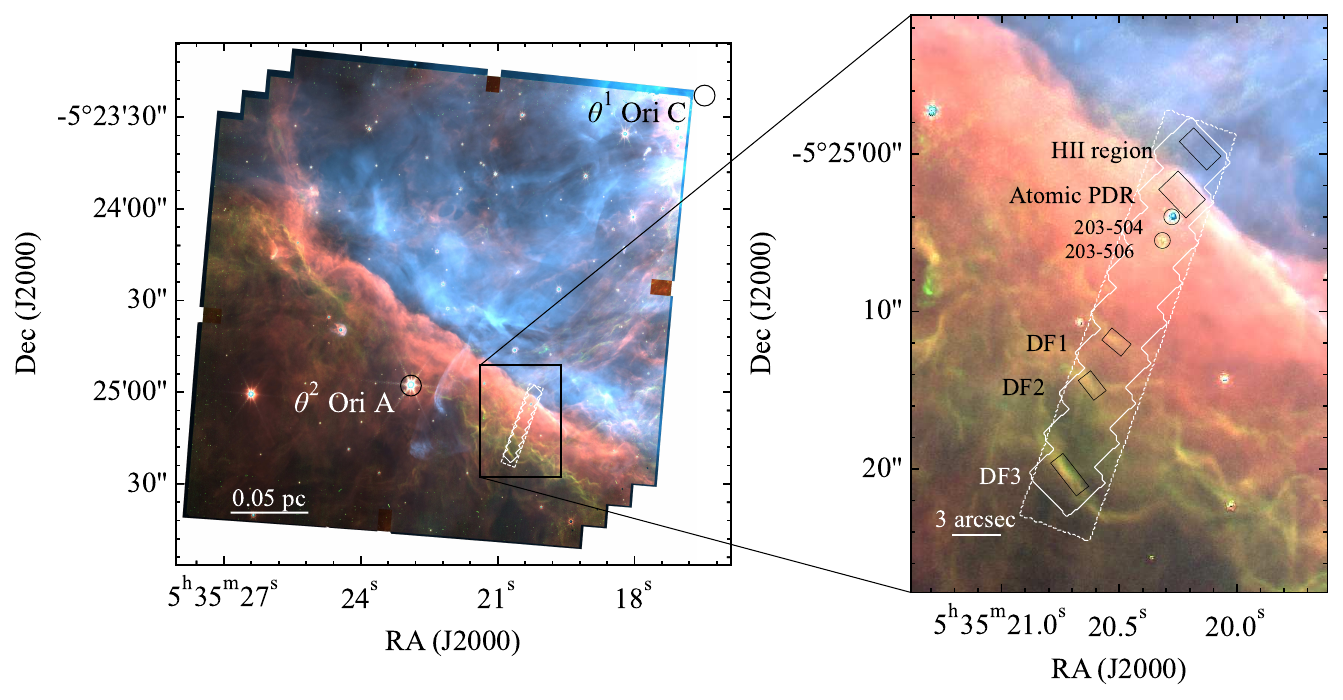}}
    \caption{PDRs4All MIRI MRS and NIRSpec footprints (dashed and solid white boundaries, respectively), and spectral extraction apertures (black boxes in the right panel) on top of a composite NIRCam image of the Orion Bar \citep[data from][]{Habart:im}. DF~1, DF~2, and DF~3 are \molh\ dissociation fronts as designated in \citet{Habart:im}. Red, green, and blue are encoded as F335M (AIB), F470N--F480M (H$_2$ emission), and F187N (Paschen $\alpha$), respectively.  %
    }
    \label{fig:apertures}
\end{figure*}

\subsection{MIRI-MRS observations and data reduction}
\label{subsec:obs}

On 30 January 2023, \jwst\ observed the Orion Bar PDR with the Mid-Infrared Instrument (MIRI) in medium resolution spectroscopy (MRS) mode \citep{wells2015, argyriou2023} as part of the PDRs4All Early Release Science program \citep{pdrs4all}. We obtained a $1\times9$ pointing mosaic in all four MRS channels (channels 1, 2, 3, and 4), and all three sub-bands within each channel (short, medium, and long). We applied a 4-point dither optimised for extended sources and use the FASTR1 readout pattern adapted for bright sources. We integrated for 521.7 s using 47 groups per integration and 4 integrations. The resulting datacube thus spans the full MRS wavelength range (4.90 to 27.90 $\mu$m) with a spectral resolution ranging from $R \sim3700$ in channel 1 to $\sim1700$ in channel 4 and a spatial resolution of 0.207\arcsec\, at short wavelengths to 0.803\arcsec\, at long wavelengths, corresponding  to 86 and 332~AU, respectively at the distance of the Orion Nebula.  

The mosaic was positioned to overlap the PDRs4All \jwst\ Near Infrared Spectrograph (NIRSpec) IFU \citep{boker2022} observations of the Orion Bar \citep[][Fig.~\ref{fig:apertures}]{Peeters:nirspec}. Given the different fields of view of the MRS channels ($\sim3$\arcsec\, in channel 1 to $\sim7$\arcsec\, in channel 4), the spatial footprint with full wavelength coverage is limited by the field of view (FOV) of channel 1. The footprint shown in Fig.~\ref{fig:apertures} represents the area with full MRS wavelength coverage, noting that the MRS data in channels 2-4 exist beyond the area shown, but we choose to use only the sub-set of data with full wavelength coverage. The NIRSpec IFU and MIRI MRS datasets combined provide a perpendicular cross-section of the Orion Bar from the \HII\ region to the  molecular zone at very high spatial and spectral resolution from 0.97 to 27.9 $\mu$m.

We reduced the MIRI-MRS data using version 1.9.5.dev10+g04688a77 of the \jwst\ pipeline\footnote{\url{https://jwst-pipeline.readthedocs.io/en/latest/}}, and \jwst\ Calibration Reference Data System\footnote{\url{https://jwst-crds.stsci.edu/}} (CRDS) context 1041. We ran the \jwst\ pipeline with default parameters except the following. The master background subtraction, outlier detection, fringe- and residual-fringe correction steps were all turned on. Cubes were built using the drizzle algorithm. The pipeline combined all pointings for each sub-band, resulting in 12 cubes (4 channels of 3 sub-bands each) covering the entire field of view.

We stitched the 12 sub-band cubes into a single cube by reprojecting all of the cubes onto a common spatial grid using channel 1 short as a reference. We then scaled the spectra to match in flux where they overlap using channel 2 long as the reference. This stitching algorithm is part of the ``Haute Couture'' algorithm described in Canin et al., (in preparation).

While the pipeline and reference files produce high-quality data products, a few artefacts still remain in the data. The most important artifacts for our analysis are fringes and flux calibration that are not yet finalised (see Appendix~\ref{sec:artefacts}). Neither of these artefacts have a strong impact on our results, as discussed in Appendix~\ref{sec:artefacts}, although we do limit our analysis to wavelengths $\leq 15~\mu$m due to the presence of artefacts beyond this range.

\subsection{Extracting template spectra from key regions}
\label{subsec:templates}

We extracted MIRI template spectra using the same extraction apertures as \citet{Peeters:nirspec}\footnote{The template spectra will be available at \url{https://pdrs4all.org}}. These apertures are selected to represent the key physical zones of the Orion Bar PDR: the \HII\ region, the atomic PDR, and the three bright \HI~/~\molh dissociation fronts (DF~1, DF~2, and DF~3) corresponding to three molecular hydrogen (\molh) filaments that were identified in the NIRSpec FOV (Fig.~\ref{fig:apertures}). We emphasize that the remaining areas in the MRS spectral map will be analysed at a later time. We note that the AIB emission detected in the \HII\ region template originates from the background PDR.  Combined with the NIRSpec templates of \citet{Peeters:nirspec}, these spectra capture all of the AIB emission in each of the five regions.  In this paper, we focus on the inventory and characterization of the AIBs found in these template spectra. We refer  to \citet{Peeters:nirspec} and \citet{lettergas} for the inventories of the gas lines from atoms and small molecules extracted from NIRSpec and MIRI MRS data, respectively. For a detailed description of the Orion Bar PDR morphology as seen by \jwst\ we refer to \citet{Habart:im} and to \citet{Peeters:nirspec}.

\subsection{Measuring the underlying continuum}
\label{subsec:continuum}

The AIB emission is perched on top of the continuum emission from stochastically heated very small grains \citep[e.g.][]{Smith:07}. Different spectral decomposition methods have deduced additional emission components referred to as: 1) emission from evaporating very small grains \citep[based on the blind signal separation method;][]{berne2007, Pilleri:12, Foschino:19} and 2) plateau emission due to large PAHs, PAH clusters and nanoparticles \citep{Bregman:89, Roche:89, Peeters:12, Boersma:14, sloan2014, peeters2017}.

In order to identify and characterize AIBs, we subtracted estimates of the continuum emission in each template spectrum. We computed a linear continuum for NIRSpec and a spline continuum anchored at selected wavelengths for MIRI data.
Furthermore, we adopted the same anchor points for all five templates. 
While our measurements of the full-width at half-maximum (FWHM) of AIBs (see Table~\ref{tab:fwhm}) do depend on the selected continuum, we note that our main goal -- to catalog AIBs and their sub-components qualitatively -- does not require highly precise estimates of the continuum. 

The FWHM\footnote{We use the terms `width' and `FWHM' interchangeably when referring to AIB profiles.} of each AIB complex was measured by normalizing the continuum-subtracted template spectrum by the peak intensity of the AIB and 
then calculating the FWHM of the entire AIB complex, that is, without taking into consideration blends, components, and/or sub-components that make up the AIB complex. The measured FWHM strongly depends on the estimated continuum emission; however, this does not impact qualitative trends in FWHM from template to template.
The integrated flux of each AIB was computed from the continuum-subtracted spectra. We refer to \citet{Peeters:nirspec} for details on how the 3.3 and 3.4 \mum~AIB fluxes were measured.

\section{AIB characteristics and assignments}
\label{sec:results}

\begin{figure*}
    \centering
    \resizebox{\hsize}{!}{
    \includegraphics[angle=90]{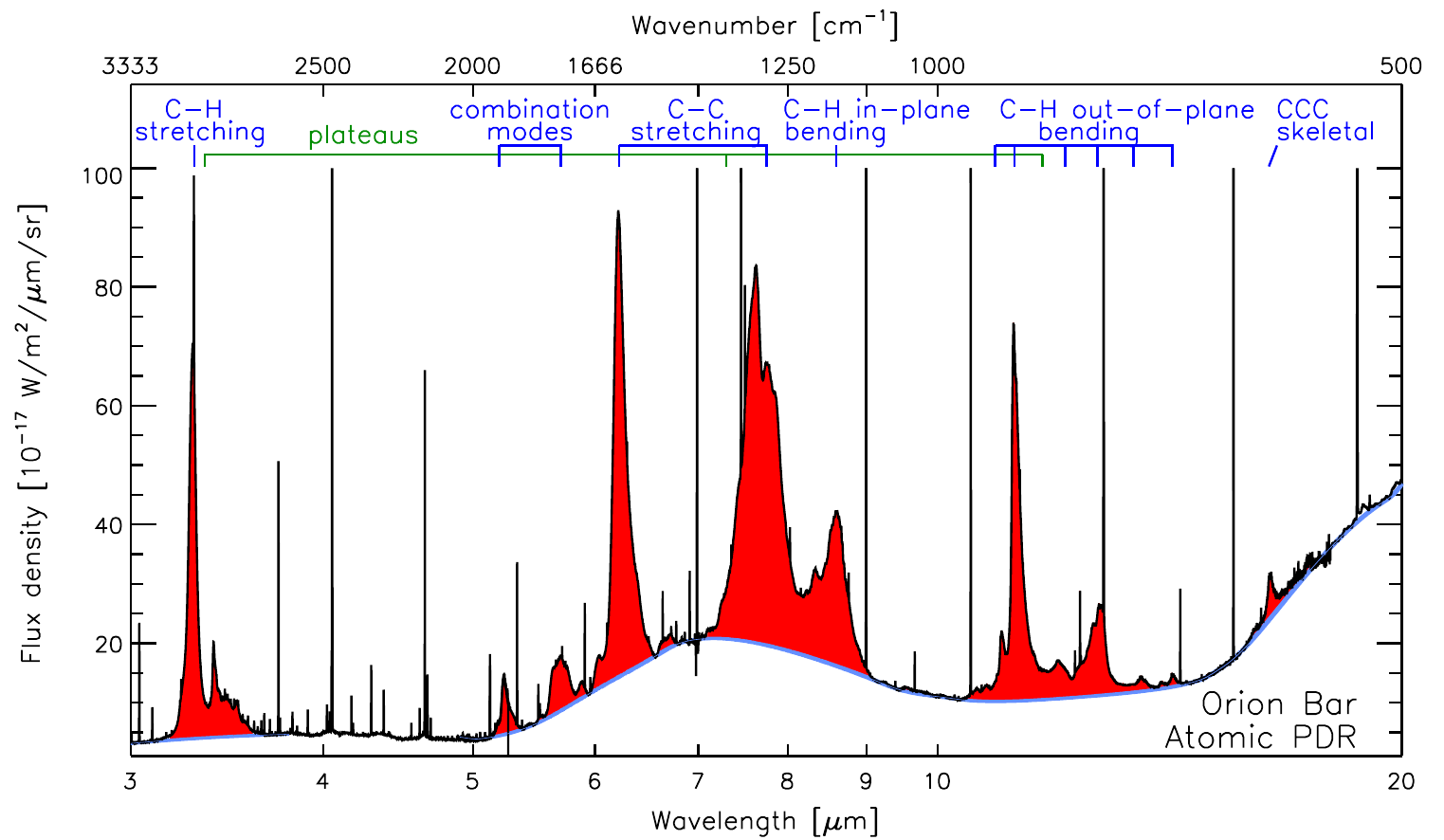}}
    \caption{AIB spectrum as seen by \jwst\ using the Orion Bar atomic PDR template spectrum (Sect.~\ref{subsec:templates}) as an example. Red shaded regions indicate emission from AIBs while blue curves indicate the underlying continuum. Figure is adapted from \citet{Peeters:colorado}.  \label{fig:classA}}
\end{figure*}

The superb quality of the Orion Bar observations combined with the increased spectral resolution compared to prior IR space observations reveals an ever-better characterization of the AIBs in terms of sub-components, multiple components making up a ``single'' band, and the precise shapes of the band profiles (see Fig.~\ref{fig:classA} for an overview and Figs.~\ref{fig:csub-aibs} and~\ref{fig:csub-aibs-zoom} for zoom-ins of selected AIBs in the five template spectra).

We offer a detailed description of the spectral characteristics of the AIB emission as seen by \jwst\ as well as current vibrational assignments in Sects.~\ref{subsec:CHstretch} to \ref{subsec:oops}. The detailed AIB inventory is listed in Table~\ref{tab:inv}. 
We note that we consider all spectrally resolved emission features to be candidate AIBs. To assess whether a candidate AIB is real or an artefact, we compare the template spectra in the location of the candidate AIB to the spectrum of the calibration standard star 10~Lac (see Appendix~\ref{sec:artefacts} for details). 
Due to the very high signal-to-noise ratio (S/N), the spectra reveal an abundance of weak features, either as standalone features, or as shoulders of other bands. Occasionally these shoulders are only visible as a change in the slope along the wing of a stronger AIB and, in such cases, we estimated the central wavelength of the weak AIB visually based on the AIB profile of the main component. 

Hereafter, all mentions of nominal AIBs, given in boldface in col. 1 of Table~\ref{tab:inv}, namely 3.3, 3.4, 5.25, 5.75, 6.2, 7.7, 8.6, 11.2, 12.0, 12.7, and 13.5 \mum, do not indicate the precise peak positions of these AIBs. The precise peak positions of these nominal AIBs are reported in terms of wavelength in col. 3 of Table~\ref{tab:inv}. We note that we converted the positions in wavelength to wavenumber by rounding to the nearest integer in units of cm$^{-1}$ and so the precision of the reported wavenumbers does not reflect the instrumental precision of the peak position of the AIBs.

\begin{figure*}
    \centering
    \resizebox{\hsize}{!}{\includegraphics{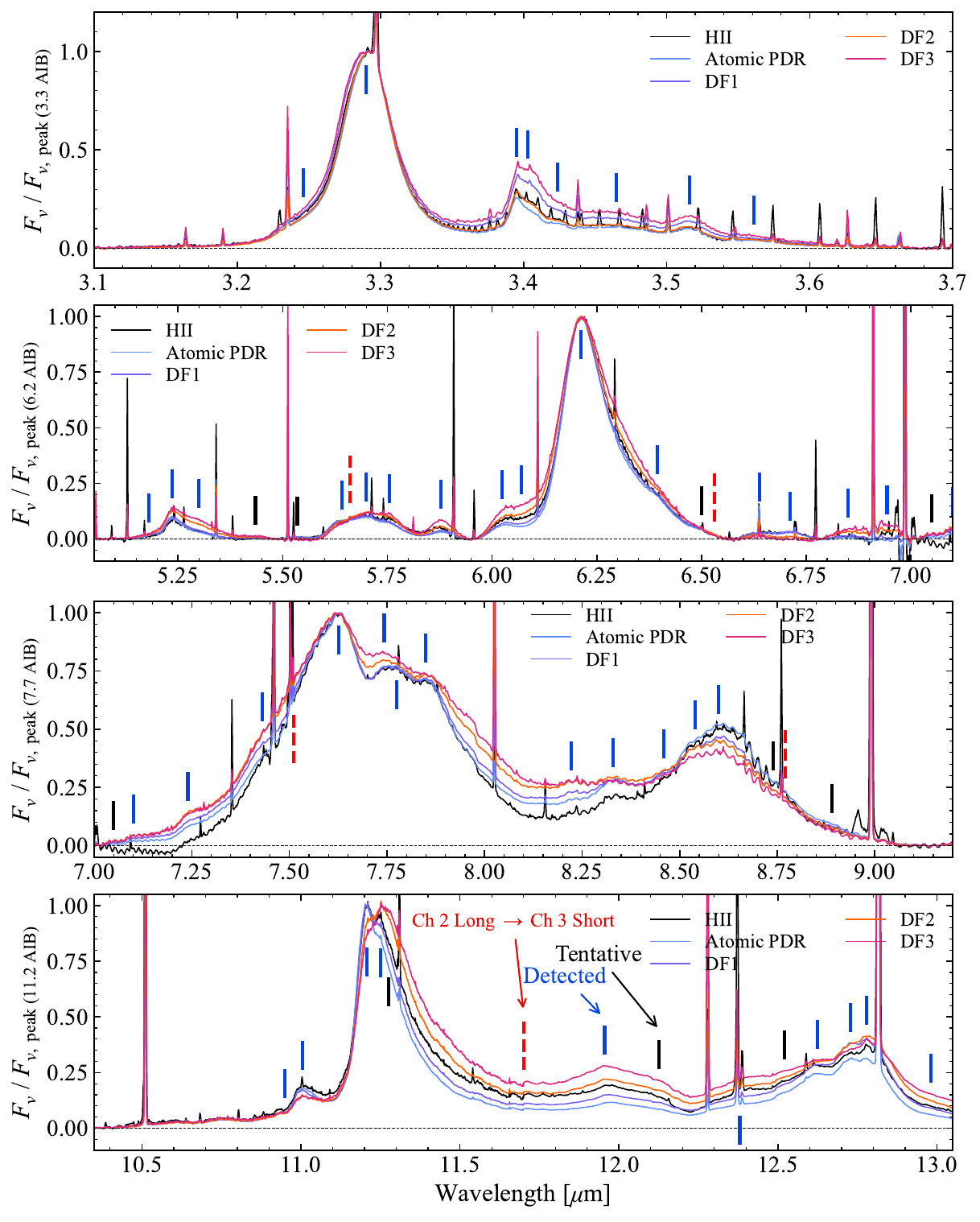}}
    \caption{Zoom-ins on the template spectra at wavelength regions centered on the 3.3~$\mu$m AIB \citep[][top]{Peeters:nirspec}, the 6.2~$\mu$m AIB (second from top), the 7.7~$\mu$m AIB (second from bottom), and the 11.2~$\mu$m AIB (bottom). Each spectrum (on an $F_\nu$ scale) is normalised by the peak surface brightness of the indicated AIB on the y-axes in each panel. The vertical tick marks indicate the positions of identified (blue) or tentative (black) AIBs and components (see Table~\ref{tab:inv} and main text). A post-pipeline correction for residual artifacts was performed for Ch2-long (10.02--11.70 \mum), Ch3-medium (13.34--15.57 \mum), and Ch3-long (15.41--17.98 \mum). For further details, see Appendix~\ref{sec:artefacts}. Red dashed vertical ticks indicate the wavelengths where we switch from using data from one MRS sub-band to another. Continued in Fig.~\ref{fig:csub-aibs-zoom}.}
    \label{fig:csub-aibs}
\end{figure*}

\begin{figure*}[!ht]
    \centering
    \resizebox{\hsize}{!}{\includegraphics{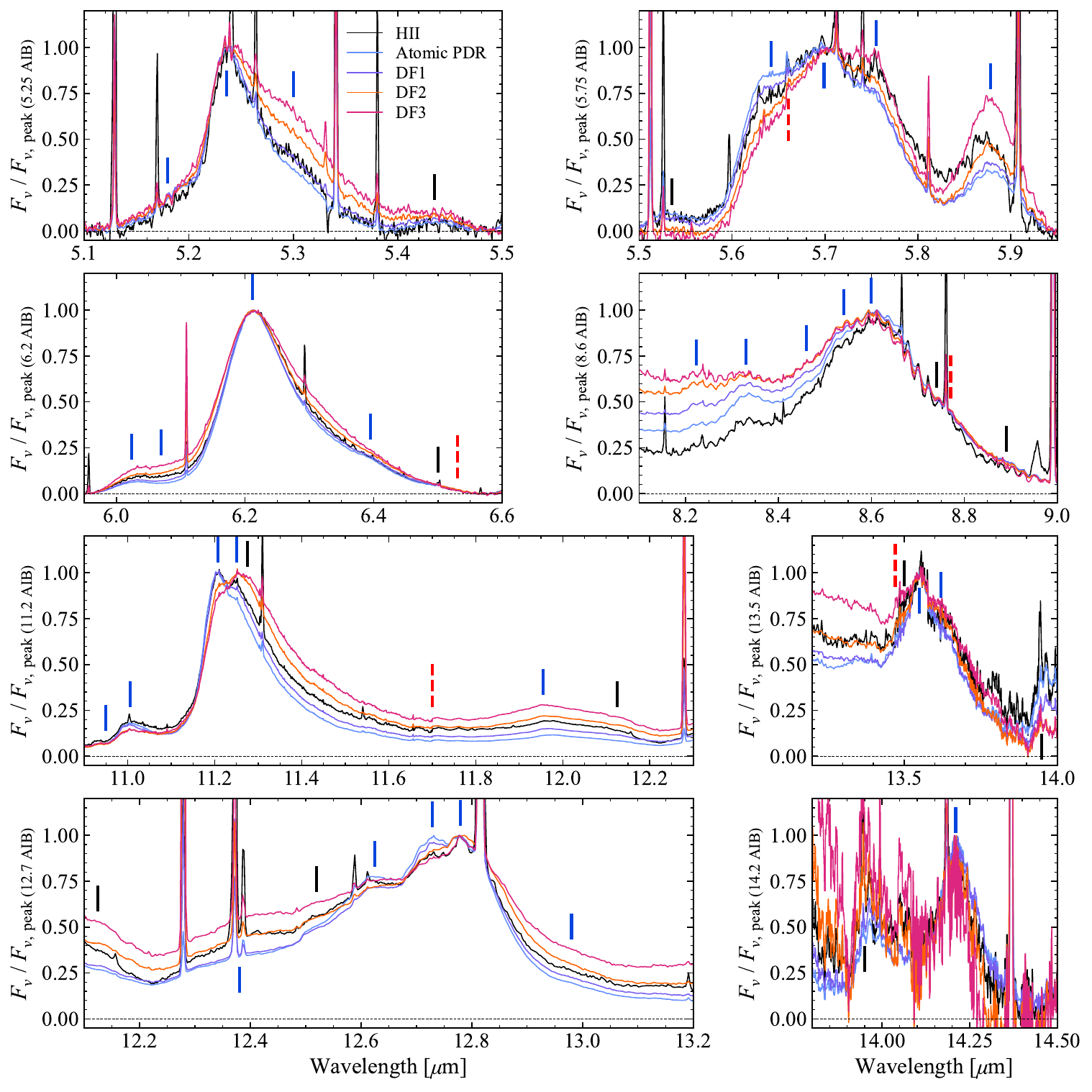}}
    \caption{Continued from Fig.~\ref{fig:csub-aibs}. From left to right, top to bottom: Zoom-ins on the template spectra normalised by the peak flux of the 5.25, 5.75 and 5.878, 6.2, 8.6, 11.2 and 12.0, 12.7, 13.5, and 14.2~\mum AIBs (indicated in the y-axis label of each panel). The panels show wavelength ranges that are also shown in Fig.~\ref{fig:csub-aibs}, except for the panels that are centered on the 13.5 and 14.2~\mum AIBs (small panels in the lower right). These figures illustrate the overall similarity and subtle differences in AIB profiles from region to region.}
    \label{fig:csub-aibs-zoom}
\end{figure*}

\begin{figure}[!ht]
    \centering
    \resizebox{\hsize}{!}{\includegraphics{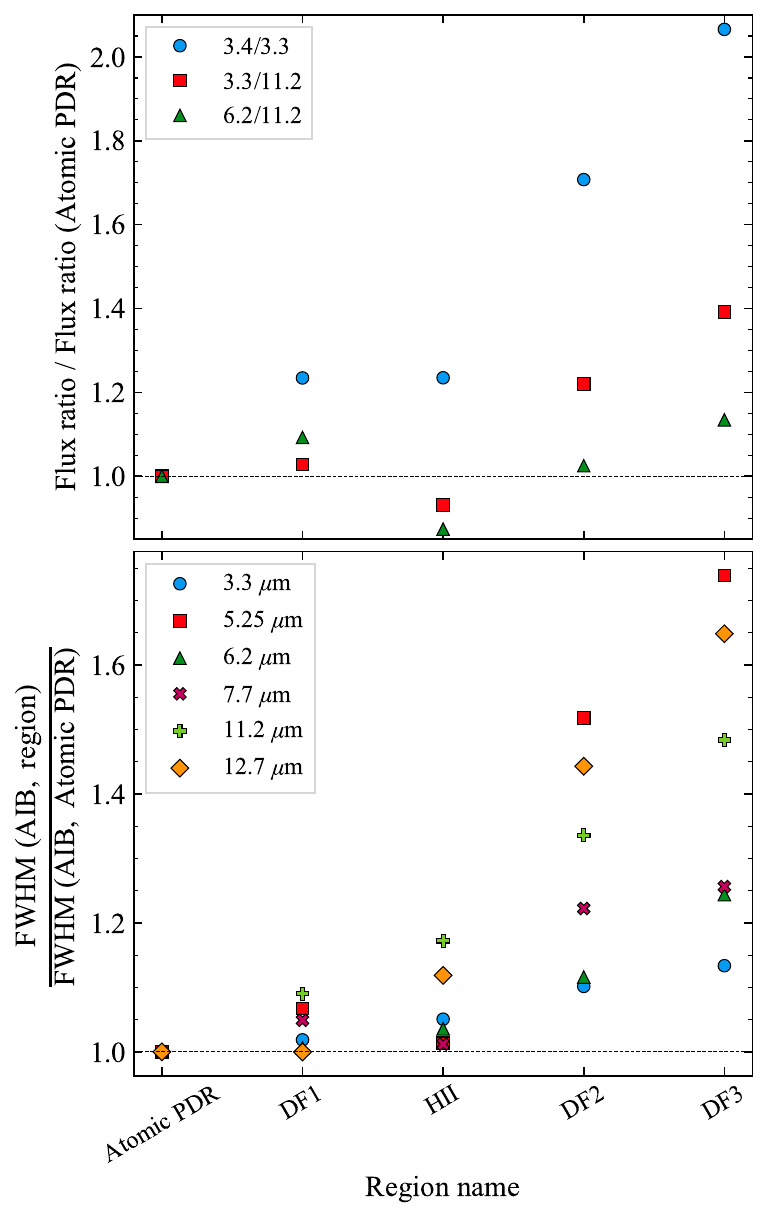}}
    \caption{Variation from region to region in the ratios of AIB integrated fluxes (top) and the widths of AIBs (bottom; data from Table~\ref{tab:fwhm}). All values are normalised to those from the atomic PDR. The ordering of the regions on the x-axis is arbitrary, but was chosen so that the data roughly show an increasing trend in FWHM from left to right (Sect.~\ref{sec:spatial}).}
    \label{fig:aib_ratio_and_fwhm_trends}
\end{figure}

\subsection{The 3.2--3.5~$\mu$m (3125--2860~cm$^{-1}$) range}\label{subsec:CHstretch}

The 3~\mum\ spectral region is dominated by the 3.3 and 3.4~\mum\ AIBs that peak  at 3.29 and 3.4 \mum, respectively. While some studies have found that the peak of the 3.3~\mum\ AIB shifts toward longer wavelengths \citep{vandiedenhoven2004}, our measurements of the peak position of this band in each template are consistent with the nominal value of 3.29~\mum.
The band profiles, however, show some slight differences among the templates. The width of the 3.29~$\mu$m band varies slightly (see Table~\ref{tab:fwhm} and Fig.~\ref{fig:aib_ratio_and_fwhm_trends}). The templates in order of increasing 3.29 $\mu$m band width are: atomic PDR, DF~1, \HII\ region, DF~2, and DF~3. While the small increase in width on the red side may be attributed to underlying broad plateau emission \citep[see][]{Peeters:nirspec}, the blue wing broadens by $\sim$5~cm$^{-1}$ on a total band width of $\sim37.5$~cm$^{-1}$ \citep[Table~\ref{tab:fwhm} and ][]{Peeters:nirspec}. 

The 3.29~\mum band is characteristic for the CH stretching mode in PAHs. The peak position is somewhat dependent on molecular structure, for example the number of adjacent hydrogens on a ring and steric hindrance between opposing hydrogens in so-called `bay' regions. Molecular symmetry has a more important effect as it controls the number of allowed transitions and the range over which IR activity is present \citep{Maltseva:2015,Maltseva:2016}. For a given PAH, the initial excitation energy has a very minor influence on the peak position \citep{mackie2022}. Earlier works have also demonstrated this minor influence \citep{Joblin:95, pech2002}. Additionally, \citet{tokunaga2021} found that the peak position and width of the 3.3~\mum feature must be fitted simultaneously as both depend on the carrier. There is also a weak dependence of peak position on the charge state, but since the CH stretch is very weak in cations \citep{Allamandola:99, peeters2002}, this is of no consequence. These modes are very much influenced by resonance effects with combination bands involving CC modes and CH in-plane bending modes \citep{Mackie:15, Mackie:16}. Overall, in the emission spectra of highly excited species, the differences in peak position mentioned here will be too subtle compared to the impact of molecular symmetry when attempting to identify the carrier(s).  The observed narrow width of the 3.3~$\mu$m AIB implies then emission by very symmetric PAHs \citep{pech2002, ricca2012, mackie2022}.

The very weak shoulder on the blue side, namely, at $\simeq 3.246$~$\mu$m, has the same strength relative to the main feature in all template spectra, suggesting it is part of the same emission complex. Its peak wavelength may point toward the stretching mode of aromatic CH groups in bay regions or, alternatively, the effect of resonant interaction in a specific species \citep{vandiedenhoven2004,candian2012,Mackie:15} or aromatic CH in polyaromatic carbon clusters \citep{Dubosq2023}.
In a recent analysis of the 3.3 \mum\ AIB in the Red Rectangle, \citet{tokunaga2022} found differences in the spectra of this source compared to earlier analyses \citep[e.g.][]{Tokunaga:33prof:91,candian2012} due to the treatment of Pfund emission lines from the standard star. \citet{candian2012} fit  the 3.3 \mum\ AIB in each spaxel of their IFU cube with two components and analysed spatial variations in the integrated intensities of these components. While issues with the standard star spectrum would affect all spectra in the cube \citep{tokunaga2022}, spatial variations in integrated intensities should not be affected.

The AIB spectra reveal a plethora of bands longward of the 3.29~$\mu$m feature between $\simeq 3.4$ and $\simeq 3.6$~$\mu$m \citep[Table~\ref{tab:inv};][]{Peeters:nirspec,sloan1997}. As the relative strengths of these sub-components show variations from source to source and within sources \citep[e.g.][]{joblin1996,Pilleri:15, Peeters:nirspec}, they are generally ascribed to different emitting groups on PAHs. Here, we note that the emission profile of the 3.4~$\mu$m band varies between the five templates, broadening to longer wavelength \citep{Peeters:nirspec}, indicating the presence of multiple components in the main 3.4~$\mu$m band at 3.395, 3.403, and 3.424~$\mu$m. The other bands do not show such profile variations. 
Bands in this wavelength range are due to the CH stretching mode in aliphatic groups and assignments to methyl (CH$_3$) groups attached to PAHs and to superhydrogenated PAHs have been proposed \citep{joblin1996,bernstein1996,Maltseva:18,buragohain2020, pla2020}. As for the aromatic CH stretching mode, the peak position is sensitive to resonances with combination bands of CC modes and CH in-plane bending modes \citep{Mackie:18}. Typically, methylated PAHs show a prominent band around 3.4~$\mu$m, but its peak position falls within a wide range, $\simeq 0.17$~$\mu$m \citep{Maltseva:18, buragohain2020}. For hydrogenated PAHs, the main activity is at slightly longer wavelengths, $\simeq 3.5$~$\mu$m within a somewhat narrower range ($\simeq 0.05$~$\mu$m). As the extra hydrogens in superhydrogenated PAHs are relatively weakly bound \citep[1.4--1.8 eV;][]{bauschlicher:14}, astronomical models imply that superhydrogenated PAHs quickly lose all these sp$^3$ hydrogens in strongly irradiated PDRs \citep[e.g. when $G_0/n(\mathrm{H})>0.03$;][]{andrews2016}. 

For further analysis of the AIB emission in this region, including the many weaker features listed in Table~\ref{tab:inv}, we refer to \citet{Peeters:nirspec}.

\begin{table}
       \begin{center}
    \caption{\label{tab:fwhm} FWHM of AIBs and AIB complexes measured from five template spectra extracted from NIRSpec-IFU (first row) and MIRI-MRS (rows 2--6) observations of the Orion Bar. }
    \small
    \begin{tabular}{lrlllllr}
    \hline\hline
     \multicolumn{2}{c}{AIB complex} & \multicolumn{1}{c}{\HII } & \multicolumn{1}{c}{Atomic}& \multicolumn{1}{c}{DF~1} & \multicolumn{1}{c}{DF~2} & \multicolumn{1}{c}{DF~3} & \multicolumn{1}{c}{$\Delta$} \\
     \multicolumn{1}{c}{} & \multicolumn{1}{c}{} &  \multicolumn{1}{c}{region} &  \multicolumn{1}{c}{PDR}  & &  & \multicolumn{1}{c}{}   \\  
     \multicolumn{1}{c}{$\mu$m} & \multicolumn{1}{c}{cm$^{-1}$} &  \multicolumn{1}{c}{} &  \multicolumn{1}{c}{}  & &  & \multicolumn{1}{c}{}   \\  
   \multicolumn{1}{c}{(1)} & \multicolumn{1}{c}{(2)} & \multicolumn{1}{c}{(3)} & \multicolumn{1}{c}{(4)} & \multicolumn{1}{c}{(5)} & \multicolumn{1}{c}{(6)} & \multicolumn{1}{c}{(7)} & \multicolumn{1}{c}{(8)}\\
    \hline 
    \multicolumn{8}{c}{}    \\[-5pt]
    3.3 & 3030 & 39.3 & 37.4 & 38.1 & 41.2 & 42.4 & 5.5\\
    5.25 & 1905 & 22.5 & 22.2 & 23.7 & 33.7 & 38.6 & 16.4\\
    6.2 & 1613 & 34.8 & 33.6 & 33.8 & 37.5 & 41.8 & 8.2\\
    7.7 & 1299 & 74.2 & 73.3 & 76.9 & 89.6 & 92.1 & 18.8\\
    11.2 & 893 & 14.3 & 12.2 & 13.3 & 16.3 & 18.1 & 5.9\\
    12.7 & 787 & 24.5 & 21.9 & 21.9 & 31.6 & 36.1 & 14.2\\
\hline
    \end{tabular}
    \tablefoot{
    The FWHM of the 8.6 \mum AIB cannot be determined, for the peak intensity of the plateau between the 7.7 \mum and 8.6 \mum AIBs is greater than half the peak intensity of the 8.6 \mum AIB (see the third panel of Fig.~\ref{fig:csub-aibs}). \\
    Columns: (1) central wavelength of AIB; (2) central wavenumber of AIB; (3)--(7) FWHM (cm$^{-1}$) of AIB complex from indicated template spectrum; (8) peak-to-peak variation in FWHM (cm$^{-1}$) observed in the templates (see Fig.~\ref{fig:aib_ratio_and_fwhm_trends}).}
    \end{center}
\end{table}

\subsection{The 5--6~$\mu$m (1600--2000~cm$^{-1}$) range}\label{subsec:overtones}

In the 5-6~$\mu$m region, previous observations have revealed two moderately weak AIB features at approximately 5.25 and 5.75~$\mu$m \citep[Table~\ref{tab:inv};][]{allamandola1989b,boersma2009}. The 5.25~$\mu$m band (Fig.~\ref{fig:csub-aibs-zoom}) consists of a broad blue shoulder centered at $\sim5.18~\mu$m and extending to about 5.205~$\mu$m, followed by a sharp blue rise to a peak at 5.236~$\mu$m and a strong red wing extending to about 5.38~$\mu$m. A detailed inspection of the profiles reveals a very weak feature at $\sim5.30~\mu$m superposed on the red wing. Comparing the five template spectra, we conclude that the 5.25~$\mu$m feature broadens, in particular on the red side, with the narrowest feature seen in the atomic PDR, and then increasing in width in the \HII\ region, DF~1, DF~2, and DF~3 (Table~\ref{tab:fwhm} and Fig.~\ref{fig:aib_ratio_and_fwhm_trends}). Besides the broadening, the observed profiles are very similar for the five templates, implying that the main feature consists of a single band. 

Inspection of the template spectra (Fig.~\ref{fig:csub-aibs-zoom}) reveals that the 5.75~$\mu$m band is a blend of three bands at 5.642, 5.699, and 5.755~$\mu$m (e.g. comparing the atomic PDR and DF~3 spectra in Fig.~\ref{fig:csub-aibs-zoom}). The MIRI MRS spectra clearly exhibit a new, symmetric feature at 5.878~$\mu$m. We also report a tentative detection of two very weak features at 5.435 and 5.535~$\mu$m.   

The spectra of PAHs show weak combination bands in this wavelength range generated by modes of the same type, for example, out-of-plane (OOP) modes \citep{boersma2009, boersma2009err}. Combination bands involving in-plane modes occur at shorter wavelengths ($3.8-4.4$~$\mu$m) and are typically an order of magnitude weaker \citep{Mackie:15, Mackie:16}. Combination bands involving the OOP bending modes typically result in a spectrum with two relatively simple AIBs near 5.25 and 5.75 \mum. For small PAHs, the ratio of the intrinsic strength of these bands to the OOP modes increases linearly 
with PAH size \citep{lemmens2019}. This ratio increases further for the larger PAHs studied in \citet{lemmens2021}. However, whether the correlation continues linearly is yet to be confirmed.

\subsection{The 6.2~$\mu$m (1610~cm$^{-1}$) AIB}\label{subsec:6.2}

The interstellar 6.2~$\mu$m band is one of the main AIBs. The profile peaks at 6.212~$\mu$m (1610~cm$^{-1}$) and has a steep blue rise, a pronounced red wing, and a blue broad shoulder centered at $\sim6.07~\mu$m. Comparing the five template spectra, we conclude that the feature broadens toward the blue side at the same time that the red wing becomes more pronounced (by about 8.2~cm$^{-1}$ on a total width of $\simeq 33.6$~cm$^{-1}$; Table~\ref{tab:fwhm}). Pending confirmation, the peak position possibly varies (its value ranges from 6.2115 to 6.2161 $\mu$m). There is a distinct weaker feature at 6.024~$\mu$m (1660~cm$^{-1}$) superposed on the blue shoulder. This symmetric feature has a constant width and
varies in intensity independently of the main feature (see Fig.~\ref{fig:csub-aibs-zoom}). This suggests that the 6.024~$\mu$m band is an independent component. We note that the observed strength variations of the 6.024~$\mu$m band do not affect the conclusion on the broadening of the blue side of the 6.2~$\mu$m band. There is a very weak feature perched on the red wing at 6.395~$\mu$m (1564~cm$^{-1}$) in the template of the atomic region. It may be obscured by the stronger red wing in the other template spectra. A very subtle change in slope of the red wing may also be present near 6.5~\mum in some templates (e.g. the atomic PDR). 

Pure aromatic CC stretching modes fall between 6.1 and 6.5~$\mu$m. In the 6--9~$\mu$m  wavelength range, the number of bands and their precise positions will depend on charge, molecular structure, size, and heterogeneity. In particular, their intrinsic strengths are very sensitive to the charge state of the species, increasing by about a factor of 10 for cations \citep{Allamandola:99, peeters2002, bauschlicher2008}. Given the observed strength of the 6.2~$\mu$m band relative to the CH stretching and OOP bending modes that dominate the neutral spectra, this AIB is attributed to PAH cations \citep{Allamandola:99}. In the past, the peak position was somewhat of an enigma. In early comparisons with harmonic calculations, this band arose at too red a wavelength in PAH cations. This problem was compounded by the adoption of a redshift of 15~cm$^{-1}$ to account for anharmonic effects during the emission cascade \citep[for a summary, see][]{bauschlicher:09}. However, model studies have revealed that anharmonicity introduces a red wing on the profile but does not lead to an appreciable redshift of the peak \citep{mackie2022}. Recent experimental and quantum chemical studies of neutral, symmetric PAHs have shown that the mismatch between the experimental and interstellar 6.2~$\mu$m band positions is less severe than thought \citep{lemmens2021}. Furthermore, quantum chemical studies on PAH cations have employed the cc-pVTZ basis set that better accounts for treatment of polarization in PAHs. With this basis set used in density functional theory calculations, the calculated peak position of the aromatic CC stretching mode in cations is in much better agreement with the observations \citep{Ricca:21}, but this still needs to be confirmed by experimental studies on PAH cations. The discrepancy noted in earlier studies between the peak position of the 6.2~$\mu$m AIB and the aromatic CC stretch in PAHs has prompted a number of suggestions. Specifically, incorporation of heteroatoms such as N into the ring backbone or coordination of atoms such as Si, the presence of aliphatic structures, protonated PAHs, and/or (pentagonal) defects will induce blue shifts in the peak position of this mode \citep{hudgins2005,pino2008,Joalland2009, carpentier2012,galue2014,Tsuge2018, WENZEL2022, rap2022}. Further studies are warranted to assess whether these suggestions are still relevant.

The observed 6.024~$\mu$m band is at too short a wavelength to be an aromatic CC stretching vibration. Rather, this position is characteristic of the C=O stretch in conjugated carbonyl groups; that is, as quinones or attached to aromatic rings \citep{allamandola1989a,sarre2019}. This band has not been the focus in many quantum chemical studies. We also note that the very weak feature at 6.395~$\mu$m is likely another aromatic CC stretching mode. 

\subsection{The 7.7~$\mu$m (1300~cm$^{-1}$) AIB complex}\label{subsec:7.7}

It has been well established that the 7.7~$\mu$m AIB is a blend of several features \citep{Bregman:hiivspn:89, Cohen:southerniras:89, peeters2002}. The \jwst\ spectra reveal that the main component at 7.626~$\mu$m is accompanied by moderately strong bands at 7.8 and 7.85~$\mu$m. The 7.8~\mum component appears narrower in DF 2 and DF 3, peaking near 7.743~\mum, although this may arise due to differences in the red wing of the 7.626~\mum component or this may reflect the lack of a different component at 7.775~\mum present in the atomic PDR, the \HII\ region and DF~1. In any case, given the observed variations between the templates, these bands are independent components. The 7.7~\mum AIB complex as a whole broadens significantly from the atomic to the molecular region (by about 18.8~cm$^{-1}$, Table~\ref{tab:fwhm}). In addition to these moderately strong components, there are also weak features at 7.24 and 7.43~$\mu$m and between the 7.7 and 8.6 complexes at 8.223 and 8.330~\mum. Very weak features are also present at shorter wavelengths (6.638, 6.711, 6.850, 6.943, 7.05, and 7.10~$\mu$m).   

Bands in this wavelength range are due to modes with a mixed character of CC stretching and CH in-plane bending vibrations. As mentioned in Sect.~\ref{subsec:6.2}, the strength of these modes is very dependent on the charge state of the species and the interstellar 7.7~$\mu$m AIB is generally attributed to PAH cations. The spectra of very symmetric PAHs become more complex with increasing size and the main band(s) in the $7.5-8.0$~$\mu$m range shift systematically with size toward longer wavelength from about 7.6 to about 7.8~$\mu$m or even larger \citep{bauschlicher2008, bauschlicher:09, ricca2012}. These quantum chemical calculations point toward compact PAHs in the size range 24 to 100 C atoms as the carriers and probably more toward the smaller size for the main 7.626~$\mu$m band and slightly larger for the two moderate components. Detailed spectral decompositions of \textit{ISO-SWS} and \textit{Spitzer-IRS} observations agree with these conclusions \citep{Joblin:08, Shannon:19}. The very weak features at 7.24 and 7.43~$\mu$m are likely also CC stretching modes. The CH deformation modes of aliphatic groups also occur around 6.8 and 7.2~$\mu$m, but these modes are weaker compared to the CH stretching modes of aliphatic groups around 3.4~$\mu$m \citep{wexler1967,yang2016, Dartois2005} and, given the weakness of the 3.4~$\mu$m AIB, we deem that identification unlikely for the weak features at 7.24 and 7.43~$\mu$m detected in all templates. The very weak features at 6.850 and 6.943~\mum are only present in DF~2 and DF~3. As both these templates also show the strongest 3.4~$\mu$m emission, these bands may arise from CH deformation modes of aliphatic groups \citep{wexler1967, Arnoult:2000}.

\subsection{The 8.6~$\mu$m (1160~cm$^{-1}$) AIB}\label{subsec:8.6}

This AIB peaks at 8.60~$\mu$m (1163~cm$^{-1}$). The apparent shift toward shorter wavelengths in the DF~3 spectrum as well as the apparent broadening of the band are likely caused by the change in the underlying ``continuum'' due to the 7.7~$\mu$m AIB and/or plateau emission and/or very small grain emission. The change in slope in the blue wing at 8.46 and 8.54~$\mu$m suggests the presence of more than one component in this AIB. However, these components seem to be very weak compared to the main band. There is a similar change in slope at 8.74~$\mu$m and potentially at 8.89~$\mu$m in all template spectra and this likely has a similar origin. Since these features are very weak, we label them as tentative.

The 8.6 \mum\ AIB is due to CH in-plane bending modes in PAHs, but this mode has a large CC stretching admixture. The intensity of this band increases significantly and it shifts to longer wavelength, producing the very prominent band that appears near 8.5~$\mu$m in the spectra of large ($N_\mathrm{C}\sim100$, $N_\mathrm{C}$ being the number of C atoms in a PAH molecule) compact PAHs \citep{bauschlicher2008}. For even larger, compact PAHs, this band starts to dominate the spectra in the $7-9$~$\mu$m range and these species are excluded as carriers of the typical AIB emission \citep{ricca2012}. In large polyaromatic and aliphatic systems, the geometrical distortions of the C-C backbone and defects, partly related to the hydrogen content, shift the position of this band \citep{carpentier2012,Dartois2020}. The weaker features on the blue side of the main band may be due to somewhat smaller and/or less symmetric PAHs while the longer wavelength feature may be due to a minor amount of somewhat larger symmetric, compact PAHs.

\subsection{The 10--20~$\mu$m (500--1000~cm$^{-1}$) range}\label{subsec:oops}

This wavelength range is dominated by the strong AIB at 11.2~\mum, a moderately strong AIB at 12.7~\mum and a plethora of weaker AIBs at 10.95, 11.005, 12.0, 13.5, 13.95, 14.21, and 16.43~\mum. The 11.2~\mum AIB clearly displays two components at 11.207 and 11.25~\mum, along with a tentative component at 11.275 $\mu$m. The AIB peaks at the first component (11.207~\mum) in the atomic PDR, the \HII\ region, and DF~1, while it peaks at the second component (11.25~\mum) in DFs 2 and 3. These two components may shift to longer wavelengths in DFs 2 and 3, however, such shifts are still yet to be confirmed. The relative strengths of these two components vary across the five templates, indicating they are independent components. The combined 11.2~\mum profile is asymmetric with a steep blue rise and a red wing. The AIB broadens significantly (by $\sim5.9$~cm$^{-1}$ on a total width of $\sim12.2$~cm$^{-1}$; Table~\ref{tab:fwhm}) through the atomic PDR, DF~1, the \HII\ region, DF~2, and DF~3 in increasing order. This broadening is driven by changes in the red wing though similar but very small changes in the steepness of the blue side are present. An additional weaker component may be present on the red wing at 11.275~\mum. In addition, similar to the 5.25~\mum AIB, the 11.2~\mum AIB displays a broad blue, slow-rising, shoulder from $\sim10.4$~\mum up to the start of the steep blue wing. A well-known weaker AIB is present at 11.005~\mum and is superposed on this blue shoulder. 

The 12.0~\mum band peaks at 11.955~\mum and may have a second component at 12.125~\mum. The template spectra furthermore display elevated emission between the red wing of the 11.2~\mum band and the 12.2~\mum band (see e.g. DF~3) suggestive of more complex AIB emission than expected based on the presence of these two bands. However, due to an artefact at 12.2~\mum (see Appendix~\ref{sec:artefacts}), confirmation of the second component, the 12.0~\mum profile, and this elevated emission between the 11.2 and 12.0~\mum bands requires further improvements to the calibration. 

The 12.7~\mum band is very complex displaying a terraced blue wing and a steep red decline. It peaks at 12.779~\mum except in the atomic PDR where it peaks at a second component at 12.729~\mum. The strengths of these components vary independently from each other. Three additional terraces are located near 12.38, 12.52, and 12.625~\mum and a red shoulder near 12.98~\mum suggests the presence of an additional component. Given the complexity of the 12.7~\mum band, spatial-spectral \jwst\ maps are required to understand its spectral decomposition into its numerous components. The entire 12.7~\mum complex significantly broadens largely on the blue side but also on the red side. We report a broadening (by $\sim14.2$~cm$^{-1}$ on a total width of $\sim21.9$~cm$^{-1}$; Table~\ref{tab:fwhm} and Fig.~\ref{fig:aib_ratio_and_fwhm_trends}) through the atomic PDR, DF~1, the \HII\ region, DF~2, and DF~3 in increasing order. 

The 13.5~\mum band peaks at 13.55~\mum and may be accompanied by two additional components at 13.50 and 13.62~\mum. The 13.5~\mum band seems to broaden as well. We note that several artefacts exist just longwards of this band (Appendix~\ref{sec:artefacts}), hampering its analysis. Hence, future improvements to the calibration and additional observations on a wider range of sources will have to confirm this broadening. These artefacts also limit the detection and analysis of bands in the 14--15~\mum range. We detected a band at 14.21~\mum and potentially at 13.95~\mum, although the latter is just to the red of the artefact at 13.92~\mum. 

Bands in the 11--14~$\mu$m range are attributed to CH OOP bending modes. The peak position and pattern of these bands is very characteristic for the molecular edge structure of the PAH; that is, the number of  adjacent H's\footnote{Some earlier comparisons of the OOP modes pattern with laboratory and quantum chemical studies included a 15 cm$^{-1}$ shift to account for anharmonicity. Recent model studies have shown that such a shift is not warranted \citep{mackie2022}.}. The bands making up the 11.2~$\mu$m AIB can be ascribed to neutral species with solo H's \citep{hony2001,bauschlicher2008}. The cationic solo H OOP band falls at slightly shorter wavelength than the corresponding solo H OOP band of neutral PAHs and the 11.0~$\mu$m AIB has been attributed to cations \citep{ Hudgins:1999, hony2001, rosenberg2011}. The 12.7 AIB complex is due to either duo H's in neutral PAHs or trio H's in cations. For species with both solo and duo H's, coupling of the duo with the solo CH OOP modes splits the former into two bands. The sub-components in the 12.7~$\mu$m AIB may reflect this coupling and/or it may be caused by contributions of more than one species with duo's.

The weak 12.0~$\mu$m AIB can be attributed to OOP modes of duo H's, while the 13.5~$\mu$m AIBs likely have an origin in OOP modes of quartet H's in pendant aromatic rings \citep{hony2001,bauschlicher2008}. The weak bands near 14.2~$\mu$m could be due to OOP modes of quintet H's. Alternatively, for larger PAHs, CCC skeletal modes are present in this wavelength range \citep{ricca2012}.

We also detected a band at 16.43 $\mu$m. Other weaker bands are present in this region but due to calibration issues (Appendix~\ref{sec:artefacts}), we refrained from characterizing them.

\section{Discussion}\label{sec:discussion}

\subsection{Comparison to previous observations}
\label{subsec:literature}

Overall, in terms of spectral inventory, the observed AIB emission in the 3~\mum range is consistent with prior high-quality ground-based observations of the Orion Bar \citep[e.g.][]{sloan1997}. Likewise, the main characteristics of the AIB emission are also detected in prior observations of the Orion Bar carried out with \textit{ISO-SWS} \citep{Verstraete:prof:01, peeters2002, vandiedenhoven2004} and \textit{Spitzer-IRS} in short-low mode \citep{Knight22}. Furthermore, in retrospect, many (weaker) bands and sub-components of the AIB emission seen by \jwst\ may also be recognised in the \textit{ISO-SWS} observation of the Orion Bar, but they were too weak and too close to the S/N limit to be reported in previous works. However, as these \jwst\ data have an unparalleled combination of extremely high S/N, spectral resolution and, most importantly, superb spatial resolution, these spectral imaging data reveal already known bands and sub-components in unprecedented detail allowing for a much improved characterization of the AIB emission. In addition, these spectral imaging data reveal previously unreported components (blends) and sub-components of the AIB emission (indicated in Table~\ref{tab:inv} and discussed in Sect.~\ref{sec:results}). 

The AIBs at 5.75, 7.7, 8.6, 11.2, and 12.7~\mum have complex sub-components. \citet{boersma2009} noted that the 5.75 AIB has an unusual profile, resembling a blended double-peaked feature. The \jwst\ template spectra indicate that the band is composed of three components with variable strengths. New components are also seen in the 12.7~\mum band. \citet{Shannon:16} reported that the 12.7~\mum band shifts to longer wavelengths at larger distances from the illuminating star in reflection nebulae. This is consistent with the behavior of this band in the Orion Bar reported here where it reflects the relative intensities of the two components at 12.729 and 12.779~\mum. These authors also reported a change in the blue wing. The \jwst\ data now characterizes the components (i.e. terraces) in the blue wing and their relative intensities. 

While the sub-components of the 8.6~\mum AIB have not, to our knowledge, been reported in the literature, several studies detail sub-components in the 7.7, 11.2, and 12.7~\mum AIBs. The 7.7~\mum AIB complex is composed of two main sub-components at $\sim$7.626 and $\sim$7.8~\mum \citep{Cohen:southerniras:89, Bregman:hiivspn:89, peeters2002}. The 7.7~\mum AIB complex is distinguished into four classes (A, B, C, and D) primarily based on its peak position \citep{peeters2002, sloan2014, matsuura:14}. 
Spectral-spatial imaging has revealed that the (class A) 7.7~\mum profile varies within extended ISM-type sources and depends on the local physical conditions: the 7.8~\mum component gains in prominence relative to the 7.626~\mum component and is accompanied by increased emission ``between'' the 7.7~\mum and 8.6~\mum AIBs in regions with less harsh radiation fields \citep{Bregman:05, berne2007, Pilleri:12, Boersma:14, peeters2017, Stock:17, Foschino:19, Knight:1333:22}. Our findings using the \jwst\ Orion Bar templates (Fig.~\ref{fig:csub-aibs}) are consistent with these past results. \citet{Pilleri:12} attributed this to an increased contribution of evaporating very small grains (eVSGs). In addition, the \jwst\ data reveal that the 7.8~\mum component is composed of three components whose relative contribution varies.

\begin{figure}
    \centering
   \resizebox{\hsize}{!}{
   \includegraphics{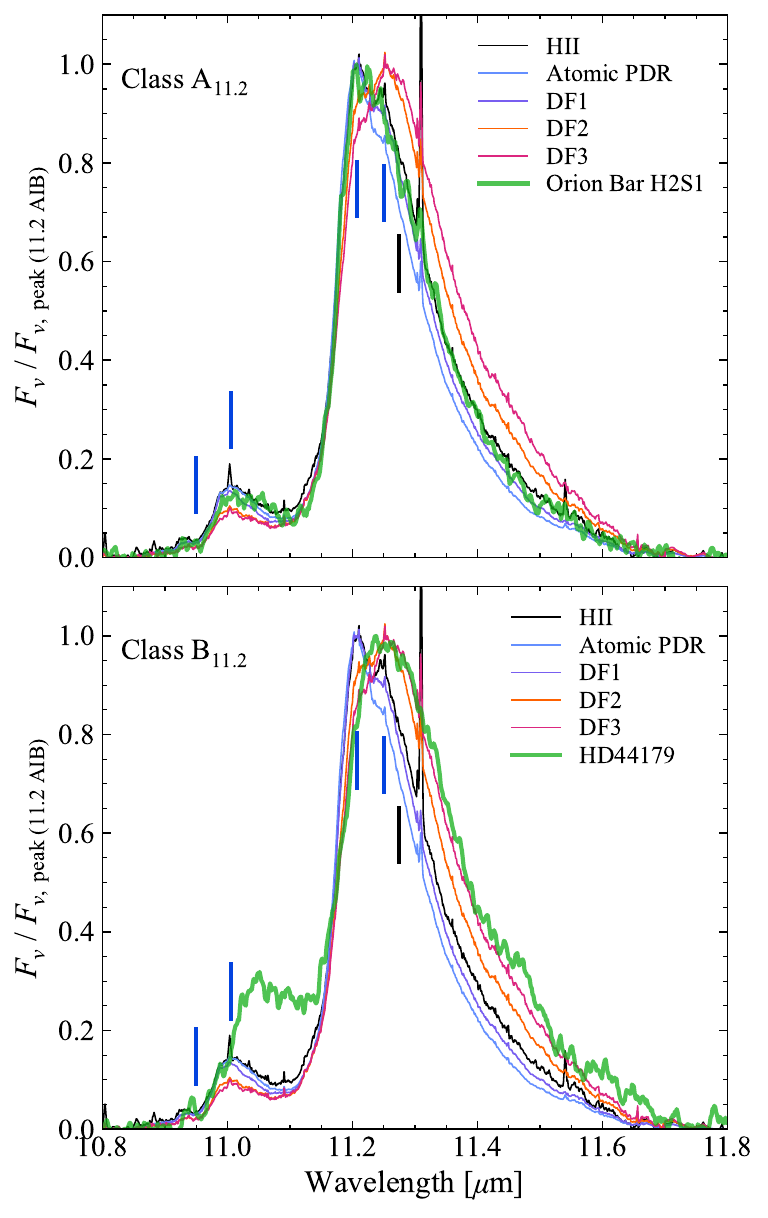}}
      \caption{Comparison of the 11.2~\mum profile in the five template spectra with a class A 11.2~\mum profile represented by the \textit{ISO-SWS} spectrum of the Orion Bar H2S1 \citep[][top panel]{vandiedenhoven2004} and a class B 11.2~\mum profile represented by the \textit{ISO-SWS} spectrum of HD~44179 \citep[][bottom panel]{vandiedenhoven2004}.  }
    \label{fig:classb}
\end{figure}

Likewise, the 11.2~\mum AIB has been classified into class A$_{11.2}$, B$_{11.2}$, and A(B)$_{11.2}$. Class A$_{11.2}$ peaks in the 11.20--11.24~\mum range and displays a less pronounced red wing relative to the peak intensity (corresponding to a FWHM of $\sim$0.17~\mum), class B$_{11.2}$ peaks at $\sim$11.25~\mum and shows a more pronounced red wing (FWHM of $\sim$0.20~\mum), and class A(B)$_{11.2}$ is a mix with a peak position as that of class A$_{11.2}$ and prominence of its red wing as that of class B$_{11.2}$ \citep[resulting in a FWHM of $\sim$0.21~\mum;][]{vandiedenhoven2004}. As for the 7.7 AIB complex, ISM-type sources display a class A$_{11.2}$ profile. Recent spectral-imaging data however indicated that the 11.2~\mum profile shifts to slightly longer wavelengths accompanied with a stronger red wing relative to the peak intensity in two (out of 17) positions of the Orion Veil \citep{Boersma:12} and in two reflection nebulae \citep{Boersma:13, Shannon:16}. These authors classified these profile variations as a shift from class A$_{11.2}$ to class A(B)$_{11.2}$, which, in the case of the two reflection nebulae, occurred when moving away from the illuminating star. \citet{Boersma:14} linked the change in the 11.2~\mum profile to a change in the 7.7~\mum AIB complex (probed by the 11.2/11.3 and 7.6/7.8 intensity ratios,  respectively). A change in peak position along with a broadening of the profile is consistent with the \jwst\ templates of the Orion Bar. As discussed in Sect.~\ref{sec:results}, the change in the peak position of the 11.2 $\mu$m AIB reflects the relative importance of two components at 11.207 and 11.25~\mum that are now clearly discerned in the \jwst\ data. Furthermore, thanks to the increased spectral and spatial resolution, we conclude that DF~3 belongs to class B$_{11.2}$ (Fig.~\ref{fig:classb}). Hence, the Orion Bar exhibits class A$_{11.2}$ profiles near the surface of the PDR which evolved from class B$_{11.2}$ profiles deeper in the molecular zone. 

The \textit{ISO-SWS} observations of the Orion Bar (taken in a 14\arcsec\,$\times$\,20\arcsec aperture) resemble the atomic PDR template, even when centered on DF~3\footnote{\textit{ISO-SWS} observation with TDT of 69501806 (uniquely identifies the ISO observation).}. This resemblance is due to the fact that the AIB emission is significantly stronger in the atomic PDR compared to the molecular PDR \citep{Habart:im, Peeters:nirspec} and it dominates the emission within the large ISO/SWS aperture. Hence, the \jwst\ spectrum of the atomic PDR in the Orion Bar (Fig.~\ref{fig:classA}) serves as the updated, high-resolution, more detailed template spectrum for class A AIB emission. The DF~2 and DF~3 templates, which probe regions deep in the molecular PDR, no longer exhibit class A$_{11.2}$ profiles while the 3.3, 6.2, 7.7, and 8.6~\mum AIBs still clearly belong to class A. A similar situation, where individual targets are found to belong to two classes, has been reported for two targets: the planetary nebula Hb~5 and the Circinus galaxy \citep{vandiedenhoven2004}. These authors furthermore found that the other two galaxies in their sample display class A profiles for the 3.3, 6.2, 7.7, and 8.6~\mum AIBs, while displaying a class A(B)$_{11.2}$ AIB profile. This suggests that, out of the main AIBs, the 11.2~\mum AIB is the cleanest indicator of the shift from class B to class A. 

\subsection{AIB profiles}\label{subsec:profiles}

Broadly speaking, the prominent AIBs can be separated into three groups: 1) bands with a steep blue rise and a pronounced red wing. The 5.25, 6.2, and 11.2~$\mu$m AIBs are clear examples. The profiles also often show a shoulder on the blue side, which is considerably weaker than the red wing; 2) bands that are clear blends of multiple components. This group includes the 3.4, 5.75, 7.7, and 12.7~$\mu$m AIBs. They typically comprise three or more sub-components.; and 3) bands that seem to be symmetric, often resembling Gaussian profiles. This group includes the 3.3, 5.878, and 8.6~$\mu$m AIBs, as well as the 6.024 and 11.005~$\mu$m AIBs. These divisions are not entirely strict. The group 1 AIBs typically have very weak features perched on the red and/or blue side of the profile. Likewise, it is conceivable that each of the blended components in the group 2 AIBs might have an intrinsic profile with relatively sharp blue rise and a more gradual red wing that is obfuscated by blending. For example, while the 3.403~$\mu$m AIB is often blended with a feature at 3.424 $\mu$m, this component is very weak and the 3.403~$\mu$m profile resembles that of group 1. We note that while the 11.2~$\mu$m AIB is also a blend of three components, the character of the profile is dominated by the presence of a steep blue rise and a pronounced red wing rather than the presence of the sub-components. Therefore, we list the 11.2~$\mu$m AIB in group 1.

Profiles with a steep blue rise and a pronounced red wing are characteristic for the effects of anharmonicity \citep{pech2002,mackie2022}. Detailed models have been developed that follow the emission cascade for a highly excited, single PAH and that include the effects of anharmonic interactions based on quantum chemical calculations \citep{mackie2022}. These models do not contain free parameters besides the size of the emitting species (i.e. the average excitation level after absorption of a FUV photon) and the resulting profiles agree qualitatively well with the observations of group 1 AIBs \citep{mackie2022}. The results show that the wavelength extent of the red wing depends on the details of the anharmonic coupling coefficients with other modes. The strength of the wing relative to the peak emission is sensitive to the excitation level of the emitting species after absorption of the UV photon (i.e. the initial average energy per mode) and the cascade process (e.g. how fast the energy is ``leaking'' away through radiative cooling). The steepness of the blue rise is controlled by rotational broadening. Analysis of the profile of the 11.2~$\mu$m AIB observed by ISO/SWS suggests emission by a modestly sized PAH \citep[$N_\mathrm{C}\sim30$;][]{mackie2022} but that conclusion has to be reassessed given the presence of more than one component in this AIB in the \jwst\ spectra of the Orion Bar.

Not all bands will show equally prominent anharmonic profiles. In particular, the far-infrared modes in small PAHs are very harmonic in nature \citep{lemmens2020} and their profiles would not develop red wings. Likewise, the aromatic CH stretching modes are very susceptible to resonances with combination bands \citep{Maltseva:2015,Maltseva:2016,Mackie:15,Mackie:16} and this interaction dominates their profiles \citep{mackie2022}.

\subsection{AIB variability in Orion}
\label{sec:spatial}

Spatial-spectral maps carry much promise to untangle the complexity of the AIBs and possibly link observed variations to the presence of specific carriers. The first forays into this field were based on \textit{Spitzer} spectral maps. Analysis of the spatial behaviour of individual AIBs and AIB components revealed their interdependence as well as new components \citep[e.g.][]{boersma2009, Peeters:12, Boersma:13, Boersma:14, Shannon:16, peeters2017}.  Analyses based on blind signal separation methods have uncovered several distinct components and spectral details \citep{berne2007, Joblin:08, Pilleri:12, Foschino:19}, but the increased spectral and spatial resolution, as well as the higher sensitivity of \jwst,\ can be expected to take this to a new level. Indeed, some of the spectral details uncovered by previous spatial-spectral studies are now directly detected in the presented \jwst\ data of the Orion Bar. Applications of blind signal separation techniques, as well as spectral fitting using PAHFIT \citep{Smith:07} and the Python PAH Database \citep{shannon-scipy2018} on the full spectral map of the Orion Bar may be a promising ground for additional detections and potential identifications. Here, we address spatial-spectral variations on inspection of the five template spectra. More detailed analyses are deferred to future studies. 

While the spectra are rich in components (Table~\ref{tab:inv}), there is little diversity between the template spectra. All templates show evidence for each sub-components, except possibly for a few very weak bands whose presence may be easily lost in the profiles of nearby strong bands. The most obvious variations are the increased prominence of the sub-components in the 7.7~$\mu$m AIB at 7.743 and 7.85~$\mu$m and in the 11.2~$\mu$m AIB at 11.25~$\mu$m in the DF 2 and DF 3 spectra, and the variation in the relative strength of the sub-components of the weak 5.75~$\mu$m band.
Similarly, the width of many AIBs (3.3, 5.25, 6.2, 7.7, 11.2, and 12.7) broadens significantly (Table \ref{tab:fwhm}). This broadening is also systematic, with the FWHM being smallest in the atomic PDR, increasing subsequently in DF~1, followed by the \HII\ region, then DF~2, and finally DF~3 (Table \ref{tab:fwhm} and Fig.~\ref{fig:aib_ratio_and_fwhm_trends}). The only exception to this systematic trend is the \HII\ region having a FWHM smaller than DF~1 (but larger than the atomic PDR) for the 5.25 and 7.7~\mum AIBs. As pointed out in Sect.~\ref{subsec:templates}, we note that the AIB emission in the \HII\ region originates from the background PDR.

There is also some region-to-region variation in the relative strengths of the main AIBs (Fig.~\ref{fig:aib_ratio_and_fwhm_trends}). The largest variations are seen in the 3.4~$\mu$m AIB to 3.3~$\mu$m AIB ratio ($\sim100$\% greater in DF~3 than in the atomic PDR) and in the 3.3~$\mu$m AIB to 11.2~$\mu$m AIB ratio ($\sim40$\% greater in DF~3 than in the atomic PDR).
Both of these ratios are largest in DF~2 and DF~3, and decrease in the atomic region. Variations in the strength of the CC modes and in-plane CH bending modes (6.2, 7.7, and 8.6~$\mu$m) relative to the 11.2~$\mu$m OOP modes are more modest (at the 10--20\% level). Larger variations occur in the relative strength of the moderate bands, namely, the 5.25, 5.878, 6.024, and 11.955~$\mu$m bands -- which are much more pronounced in DF 2 and DF 3 than in the atomic zone. As noted in Sect.~\ref{subsec:literature}, the \textit{ISO-SWS} data of the Orion Bar resembles the atomic PDR template. Hence, the range in spectral variability within class A AIBs is well represented by the five templates for all AIBs except the 11.2~\mum AIB. For the 11.2~\mum AIB, the presented data not only showcase the class A AIB variability but also the shifts from class A to class A(B) and then to class B. It is expected that future \jwst\ observations probing a large range of physical conditions and environments further extend the spectral variability in the AIB emission.

The anharmonic profile of bands due to smaller PAHs will tend to have a less steep blue rise due to the increase in the rotational broadening as well as a slight increase in the width and a more pronounced red wing due the higher internal excitation for the same photon energy \citep{mackie2022,tielens2021}. Hence, the presence of  somewhat smaller PAHs may be at the origin (of some) of the overall profile variations in the 3.4, 5.25, 6.2, and 11.2~$\mu$m AIBs in the \jwst\ template spectra. 
We note that the spectrum of the DF~1 template resembles much more the atomic region template than the DF~2 and DF~3 dissociation front templates (Figs.~\ref{fig:csub-aibs} and~\ref{fig:csub-aibs-zoom}, as well as Table~\ref{tab:fwhm}). This is likely due to the terraced-field-like structure of the molecular PDR resulting in a strongly enhanced line-of-sight visual extinction through the foreground atomic PDR toward DF~1 compared to DF~2 and DF~3 \citep{Habart:im, Peeters:nirspec}. Hence, a large contribution from the atomic region in the foreground is contributing to the emission toward DF~1. 
The overall similarity of the spectra suggests that the PAH family is very robust but has a small amount of additional species in the DF 2 and DF 3 zones that is not present in the surface layers of the PDR. In the Orion Bar, the PDR material is advected from the molecular zone to the ionization front at about 1~km s$^{-1}$ \citep{pabst2019} over $\simeq20,000$~yr. In that period, a PAH will have absorbed some $10^{8}$ UV photons and, yet, apparently the effect on the composition of the interstellar PAH family is only minor as it only results in a change in the prominence of the sub-components in the 7.7 and 11.2 $\mu$m AIBs. This likely reflects that for moderate-to-large PAHs ($N_\mathrm{C}\gtrsim30$), photofragmentation is a minor channel compared to IR emission and, moreover, when fragmentation occurs, the H-loss channel dominates over C loss \citep{Allain1996a, Allain1996b, zhen2014a, zhen2014b, Wenzel2020} and then rapidly followed by rehydrogenation with abundant atomic H \citep{Montillaud2013, andrews2016}. We note that the UV field increases by about two orders of magnitude between the H$_2$ dissociation front and the PDR surface but the atomic H abundance increases by a similar factor as H$_2$ is increasingly photolysed near the surface. Hence, the ratio of the local FUV field to the atomic hydrogen density, $G_{0}/n(\mathrm{H})$, which controls the photoprocessing \citep{andrews2016}, does not vary much among the five template regions. We thus suggest that the additional species in the deeper layers of the Orion Bar causing the increased prominence of sub-components in the 5.75 (at 5.755~\mum), 7.7, and 11.2 $\mu$m AIBs, are aromatic species and/or functional groups and/or pendant rings that are more susceptible to photolysis.   

Photolytic processing of the PAH family with position in the Orion Bar may also leave its imprint on the PAH size distribution and this will affect the relative strength of the AIBs. The 3.3/11.2 AIB ratio has long been used as an indicator of the size of the emitting species \citep{allamandola1989a,pech2002,ricca2012,Mori:2012,croiset2016,Maragkoudakis:20, Knight21, Knight22} as this ratio is controlled by the `excitation temperature' of the emitting species and, hence, for a fixed FUV photon energy, by the size. For the five \jwst\ template regions, the 3.3/11.2 AIB ratio is observed to vary by about 40\%, being largest in DF~3 and decreasing toward DF~2, followed by DF~1, the atomic PDR and the \HII\ region. The variation in this ratio is slightly less than what is observed in the reflection nebulae NGC~7023 and in the larger Orion region (e.g. the Orion Bar and the Veil region beyond the Orion Bar) and corresponds to an increase in the typical size of the emitting species by about 40\%\ toward the surface \citep{croiset2016,Knight21, Knight22, murga2022}. Hence, we link the decreased prominence of the sub-component in the 5.75~\mum (at 5.755~\mum), 7.7~\mum (at 7.743 and 7.775~$\mu$m), and 11.2~\mum AIBs (at 11.275~$\mu$m) as well as the variation in the 3.3/11.2 AIB ratio to the effects of photolysis as material is advected from the deeper layers of the PDR to the surface. 

Variations in the 6.2/11.2 ratio (Fig.~\ref{fig:aib_ratio_and_fwhm_trends}) are generally attributed to variations in the ionised fraction of PAH \citep{peeters2002,galliano2008,stock2016,boersma2018}. This ratio is only 12\%\ stronger in DF~3 than the surface of the PDR. The limited variations in the 6.2/11.2 AIB ratio is at odds with those measured by \textit{Spitzer} and \textit{ISO}. Specifically, this ratio is observed to increase by about 50\%\ across the Orion Bar when approaching the Trapezium cluster \citep{Knight22}. Moreover, \citet{galliano2008} measured an increase in this ratio by almost a factor of 2 over (a much wider swath of) the Orion Bar. The PAH ionization balance is controlled by the ionization parameter, $G_0T^{1/2}/n_e$ with $G_0$, $T$, and $n_e$ the intensity of the FUV field, the gas temperature, and the electron density. The high spatial resolution of \jwst\ allows for a clear separation of the emission at the H$_2$ dissociation fronts and the PDR surface. The limited variation in the 6.2/11.2 AIB ratio is somewhat surprising because the PAH ionizing photon flux differs by about a factor of 40 between the dissociation fronts (located at $A_V=2$ mag) and the PDR surface; the gas temperature will also increase (slightly) toward the surface, while the electron abundance remains constant over this region  \citep{TH1985}. It is also possible that PAH cations contribute an appreciable amount to the 11.2~$\mu$m band \citep{Shannon:16, boersma2018}. Further modelling will be important to fully understand the complexity of the Orion Bar (A. Sidhu et al., in prep.).

The 3.4/3.3 AIB ratio is observed to decrease by about 100\% from DF~3 to the atomic PDR (Fig.~\ref{fig:aib_ratio_and_fwhm_trends}). Such variations have also been seen in other nebulae and attributed to photofragmentation processes \citep{joblin1996}. The 3.4~$\mu$m AIB is attributed to aliphatic CH modes in the form of a minor amount of H bonded to sp$^3$ C atoms either in the form of methyl groups or as superhydrogenated PAHs \citep{schutte1993,bernstein1996,joblin1996}. The abundance of superhydrogenated PAHs is expected to be very small throughout the Orion Bar as such extra H's are readily lost in the strong FUV radiation field \citep{andrews2016}. Methyl groups are also more easily photolysed than aromatic H's \citep[energy barriers are 3.69, 4.00, and 4.47 eV for CH$_2$-H, -CH$_3$ and aromatic H loss, respectively;][]{tielens2021}. Recent experiments report that, for cations, this methyl group photolysis can lead to quite stable tropylium formation \citep[loss of H followed by isomerization to a seven-membered ring; ][]{Jochims:99, Zhen2016,WENZEL2022}. However, further investigation is required to firmly establish the importance of this fragmentation route for conditions present in the interstellar medium. If borne out, the reaction of the tropylium cation with atomic H has a calculated barrier of 3.2 kcal mol$^{-1}$ \citep[1600~K;][]{bullins2009} and, hence, under warm, dense H-rich conditions, the methyl group could be reformed. Hence, in a ``suitable'' PDR,  the species may cycle back and forth between a methyl functional group and the tropylium structure until eventually -CH$_3$ loss occurs. In any case, it can be expected that the stronger UV field nearer to the surface will reduce the number of CH methyl groups compared to the number of CH aromatic  bonds. Further experimental and quantum chemical studies will have to address the competition between the various channels involved in the chemistry of methylated PAHs in PDRs.

The increased importance of PAH photolysis near the surface of the Orion Bar PDR is in line with the GrandPAH hypothesis that only the most resilient species can sustain the harsh conditions of  strong FUV radiation fields. Thus, a limited number of compact, large PAHs will dominate the interstellar PAH family in these conditions \citep{andrews2015,tielens2013}. If the conditions are right, large PAHs may even be stripped of all their H's and isomerize to the fullerene, C$_{60}$ \citep{Boersma:12,berne2015,zhen2014b}. We also realize that the presence of somewhat smaller PAHs and the increased importance of aliphatic functional groups deep in the PDR may reflect the importance of ion-molecule and/or radical chemistry during the preceding dark cloud core phase modifying and/or forming PAHs in a bottom-up scenario akin to that proposed for the formation of benzonitrile, indene, and cyanonaphthalene \citep{McGuire2018,Cernicharo2021,mcguire2021}.

\section{Conclusions}
\label{sec:conclusions}

The superb sensitivity and spectral resolution %
of \jwst\  have revealed an ever-better characterization of the AIBs in the Orion Bar in terms of sub-components, multiple components making up a ``single'' band, and band profiles.  In addition, the unprecedented spatial resolution of the spectral imaging data showcases the interdependence of the numerous AIB components. 

We extracted five template spectra in apertures positioned on the \HII\ region, the atomic PDR, and the three dissociation fronts DFs~1, 2, and 3. The spectra display a wealth of detail and many weak features have now been firmly identified, their peak positions quantified, and their profiles established. At the same time, the spectra are really very simple. There are a limited number of strong bands with well defined peak positions and red-shaded profiles characteristic for anharmonic interactions. And 
there is little diversity between the templates. A modest variation is observed in the relative intensities of the main AIBs and of the sub-components of an AIB as well as a systematic broadening of the FWHM of many AIBs (smallest in the atomic PDR and largest in DF~3). Consequently, these templates demonstrate the spectral variations in the class A AIB emission as well as the shift from class B$_{11.2}$ (DF~3) to class A$_{11.2}$ (atomic PDR). The comparison of the template spectra with the \textit{ISO-SWS} spectrum of the Orion Bar underscores that the spectrum of the atomic region is the "poster child" for the class A spectrum \citep[Fig.~\ref{fig:classA};][]{peeters2002, vandiedenhoven2004}. This comparison also demonstrates that in a large aperture, PDRs such as Orion are expected to show class A spectra. Conversely, PDRs with more gentle physical conditions (e.g. in the DF~3) are expected to display a slightly modified class A AIB spectrum (except for the 11.2~\mum AIB), showcasing broader AIBs and an increased prominence of minor sub-components with respect to the exemplar class A spectrum. In the case of the 11.2~\mum AIB, more gentle physical conditions broaden the AIB and increase the prominence of minor sub-components seen in class A, resulting in a class B 11.2~\mum AIB profile. Hence, the templates suggest a shift from class B$_{11.2}$ (DF~3) to class A$_{11.2}$ (atomic PDR). Further modelling of the PDR physics and chemistry may help to pinpoint the physical and chemical processes that drive these spatial-spectral variations in the Orion Bar. Furthermore, we expect that similar studies of a variety of sources with \jwst\ will provide deeper insight in the origin of the A, B, C, and D classes identified by using measurements obtained with \textit{ISO} and \textit{Spitzer} \citep{peeters2002, vandiedenhoven2004, sloan2014, matsuura:14}. In any case, the spatial-spectral variations in the Orion Bar  provide a framework in which AIB spectra of extragalactic regions of massive star formation can be analysed in terms of the physical conditions in their PDRs.

An analysis of the \textit{Spitzer} spectra of a variety of objects revealed that the mid-IR spectra at the brightest spots in PDRs show remarkably similar AIBs and this has been taken to imply that in the harsh conditions of these positions, the PAH family is dominated by a few species that can withstand the harsh conditions in PDRs \citep{andrews2015,tielens2013}. For now, the limited diversity in the AIB characteristics of the Orion Bar templates points in the same direction. Indeed, a broad distribution of PAHs would result in  much more sub-structure and variation behavior. In addition, the disappearance of the (weak) 11.25~\mum component of the 11.2~\mum AIB in the PDR surface layers implies that photochemistry is important: only the most robust species survive in the harsh conditions at the surface of the PDR. We note that while the cosmic AIB emission can be classified in four classes (A, B, C, and D), interstellar AIB emission invariably belongs to class A. The Orion Bar spectrum is the "poster child" of the class A spectrum -- out of all the AIBs found in the Orion Bar template spectra, only the 11.2~\mum AIB shows some indication of a class B contribution. Only very distinctly different classes of objects with unique histories display classes B, C, and D AIB emission. This too implies that the interstellar PAH family consists of a small set of very robust species.

Moreover, we conclude that the profiles of the 5.25, 6.2, and  11.2 $\mu$m AIBs are controlled by anharmonicity, rather than by blending of a number of bands, while variations in the widths of these bands in the different template spectra are related to variations in the excitation of the emitting species in those positions. As a corollary, this implies that these bands are likely dominated by emission of a single carrier, further supporting the GrandPAH hypothesis \citep{mackie2022}. However, it remains to be seen whether this spectral similarity still holds when a much larger range of objects is investigated at the higher spectral resolution of \jwst.

The much higher spatial resolution of \jwst\ provides further insight in the processes that might be relevant for the composition of the PAH family. Specifically, as argued in Sect.~\ref{sec:discussion}, the decreased prominence of the minor features in the 5.75, 7.7, and 11.2 $\mu$m AIBs in the atomic region indicates the loss of a sub-population of the PAH family \citep{berne2012, Montillaud2013, berne2015, andrews2016} or the loss of very small grains \citep{Pilleri:12, Pilleri:15} in this region.  Likewise, the decrease in the methyl group coverage, as evidenced in the variation of the 3.4/3.3 AIB ratio, indicates the loss of more loosely bound functional groups in the surface layers or their conversion to aromatic moieties. Models suggest that photolysis of PAHs is controlled by the strength of the UV field over the atomic hydrogen density \citep{Montillaud2013,andrews2016,berne2015} and over much of the Orion Bar PDR, $G_0/n(\mathrm{H})\simeq 1$ \citep{TH1985,jeronimo2012}, implying that the PAHs that are lost from the PAH family are small ($N_\mathrm{C} \lesssim 50$). 

Hence, the picture that emerges from the analysis of the template spectra is the increased importance of FUV processing in PDR surface layers, resulting in a ``weeding out'' of the weakest links of the PAH family. These less resistant species are possibly formed from small hydrocarbons during the preceding dark cloud phase of the region in a bottom-up chemical scenario \citep{Cuadrado:15, mcguire2021}, feeding the gas with small hydrogen rich photolytically produced species \citep{Alata2015}. The UV processing of the PAH family will start with the loss of the smallest PAHs but as the PDR material is advected to the surface, larger and larger PAHs will become susceptible to photoprocessing. Similar scenarios have been proposed recently with data not obtained with \jwst, albeit with slightly different numbers \citep[e.g. $N_\mathrm{C}\lesssim50$ in][]{murga2022}. In favourable conditions, the processing of very large PAHs ($N_\mathrm{C}\gtrsim60$) may lead to the formation of C$_{60}$ \citep{berne2012,berne2015}. While the present data around 18.9 $\mu$m are still marred by instrumental artefacts, future searches for the signature of fullerenes in the MIRI spectra of these surface layers may reveal whether such a scenario plays a role under the conditions of the Orion Bar. 

\jwst\ is poised to obtain high-quality spectral imaging observations of a large sample of environments probing the full range of physical conditions that are relevant for AIB emission. These observations promise to capture the complexity of AIB emission with the unprecedented detail that is needed in order to advance our understanding of the photochemical evolution of large carbonaceous molecules. 

\begin{acknowledgements}
We thank the referee, Alan Tokunaga, for constructive comments on the manuscript.
We are very grateful to the \jwst\ Help Desk for their support with pipeline and calibration issues that were encountered while writing this paper. 
This work is based on observations made with the NASA/ESA/CSA James Webb Space Telescope. The data were obtained from the Mikulski Archive for Space Telescopes at the Space Telescope Science Institute, which is operated by the Association of Universities for Research in Astronomy, Inc., under NASA contract NAS 5-03127 for JWST. These observations are associated with program \#1288.
Support for program \#1288 was provided by NASA through a grant from the Space Telescope Science Institute, which is operated by the Association of Universities for Research in Astronomy, Inc., under NASA contract NAS 5-03127.
EP and JC acknowledge support from the University of Western Ontario, the Institute for Earth and Space Exploration, the Canadian Space Agency (CSA; 22JWGO1- 16), and the Natural Sciences and Engineering Research Council of Canada.
Studies of interstellar PAHs at Leiden Observatory (AT) are supported by a Spinoza premie from the Dutch Science Agency, NWO. 
CB is grateful for an appointment at NASA Ames Research Center through the San Jos\'e State University Research Foundation (80NSSC22M0107). 
Part of this work was supported by the Programme National ``Physique et Chimie du Milieu Interstellaire'' (PCMI) of CNRS/INSU with INC/INP co-funded by CEA and CNES. 
JRG and SC thank the Spanish MCINN for funding support under grant \mbox{PID2019-106110GB-I00}. 
TO is supported by JSPS Bilateral Program, Grant Number 120219939. 
Work by YO and MR is carried out within the Collaborative Research Centre 956, sub-project C1, funded by the Deutsche Forschungsgemeinschaft (DFG) – project ID 184018867. This work was also partly supported by the Spanish program Unidad de Excelencia María de Maeztu CEX2020-001058-M, financed by MCIN/AEI/10.13039/501100011033.
NN is funded by the United Arab Emirates University (UAEU) through UAEU Program for Advanced Research (UPAR) grant G00003479.
AP (Amit Pathak) would like to acknowledge financial support from Department of Science and Technology - SERB via Core Research Grant (DST-CRG) grant (SERB-CRG/2021/000907), Institutes of Eminence (IoE) incentive grant, BHU (incentive/2021- 22/32439), Banaras Hindu University, Varanasi and thanks the Inter-University Centre for Astronomy and Astrophysics, Pune for associateship.
This work is sponsored (in part) by the CAS, through a grant to the CAS South America Center for Astronomy (CASSACA) in Santiago, Chile.
AR gratefully acknowledges support from the directed Work Package at NASA Ames titled: ‘Laboratory Astrophysics –The NASA Ames PAH IR Spectroscopic Database’.
MB acknowledges DST for the DST INSPIRE Faculty fellowship.
HZ acknowledges support from the Swedish Research Council (contract No 2020-03437).
P.M. acknowledges grants EUR2021-122006, TED2021-129416A-I00 and PID2021-125309OA-I00 funded by MCIN/AEI/ 10.13039/501100011033 and European Union NextGenerationEU/PRTR.
MSK is funded by RSCF, grant number 21-12-00373.

\end{acknowledgements}

\bibliographystyle{aa}
\bibliography{46662corr}

\begin{appendix}

\section{Inventory of AIBs}
\label{sec:inventory}

Table~\ref{tab:inv} shows the inventory of AIBs identified in this study.

\clearpage
\onecolumn
\begin{longtable}{llllllp{3.2cm}p{5.3cm}} 
    \caption{Catalogue of AIB features detected in the five template spectra of the Orion Bar. The method used to select these features is described in Sect.~\ref{sec:results}. The precise peak positions of nominal AIBs -- whose wavelengths appear in boldface in this table -- are indicated in Col. 3. We note that we converted the positions in wavelength to wavenumber by rounding to the nearest integer in units of cm$^{-1}$, and so the precision of the reported wavenumbers do not reflect the precision of the peak position of the AIBs.}\label{tab:inv} \\
\hline\\[-8pt]
    \multicolumn{2}{c}{AIB} & \multicolumn{2}{c}{Components} &  New?  & $I_\mathrm{peak}$ & Characteristics & Assignment \\
    \multicolumn{1}{c}{$\mu$m} & \multicolumn{1}{c}{cm$^{-1}$} &  \multicolumn{1}{c}{$\mu$m} & \multicolumn{1}{c}{cm$^{-1}$} & &  &    \\    
    \multicolumn{1}{l}{(1)} & \multicolumn{1}{l}{(2)} & \multicolumn{1}{l}{(3)} & \multicolumn{1}{l}{(4)} & \multicolumn{1}{l}{(5)} & \multicolumn{1}{l}{(6)} & \multicolumn{1}{l}{(7)} & \multicolumn{1}{l}{(8)}\\
    \\[-8pt]
    \hline 
    \endfirsthead
    \caption{continued.}\\
    \hline\\[-8pt]
    \multicolumn{2}{c}{AIB} & \multicolumn{2}{c}{Components} & New? & $I_\mathrm{peak}$ & Characteristics & Assignment \\
    \multicolumn{1}{c}{$\mu$m} & \multicolumn{1}{c}{cm$^{-1}$} &  \multicolumn{1}{c}{$\mu$m} & \multicolumn{1}{c}{cm$^{-1}$} & &  &    \\    
    \multicolumn{1}{c}{(1)} & \multicolumn{1}{c}{(2)} & \multicolumn{1}{c}{(3)} & \multicolumn{1}{c}{(4)} & \multicolumn{1}{c}{(5)} & \multicolumn{1}{c}{(6)} & \multicolumn{1}{c}{(7)} & \multicolumn{1}{c}{(8)}\\
    \\[-8pt]
    \hline
\endhead
\hline
\endfoot
& \\[-5pt]
\multicolumn{8}{c}{\textbf{3~$\mu$m (3000~cm$^{-1}$) region\tablefootmark{a}}}\\
\textbf{3.3} & 3030 & & &  & vs & & (aromatic) CH stretches\\
& & 3.246 & 3081&  & ms & symm, blend &  \\
& & 3.290 & 3040&  & vs & symm, blend  & \\
\textbf{3.4} & 2941& & &  & s, ms & asymm, blend & CH stretches in aliphatic groups (methyl (CH$_3$) and ethyl (CH$_2$CH$_3$) groups attached to PAHs and   superhydrogenated PAHs) \\
& &  3.395 & 2946&  & s, ms & blend &  \\
& &  3.403 & 2939&  & s, ms & blend &  \\
& &  3.424 & 2921& \checkmark & ms, mw &  blend &  \\ 
3.465 & 2886& & &  & ms, mw & symm, blend & CH stretches in aliphatic groups (see 3.4~\mum AIB) \\
3.516 & 2844& & &  & ms, mw & symm, blend & CH stretches in aliphatic groups (see 3.4~\mum AIB) \\ 
3.561 & 2808& & &  & mw, w & symm, blend & CH stretches in aliphatic groups (see 3.4~\mum AIB)  \\
\\[-5pt]
\arrayrulecolor{lightgrey}\hline
 & \\[-5pt]
\multicolumn{8}{c}{\textbf{5--6~$\mu$m (1600--2000~cm$^{-1}$) range} }\\[5pt]
\textbf{5.25} & 1905 & & & & mw & asymm & combination bands generated by modes of the same type, e.g., out-of-plane (OOP) modes \\
& & $\sim$5.18 & $\sim$1931& & w, vw & blue shoulder&  \\
& & 5.236 & 1910&  & mw & asymm &   \\
& & $\sim$5.30 & $\sim$1887& \checkmark & vw\tablefootmark{b} & blend &     \\ 
5.435? & 1840?& & & \checkmark & vw & symm?, often blend (5.25) & combination bands (see 5.25~\mum AIB)\\
5.535? & 1807?& & &  \checkmark& vw & symm?, blend (5.75) &combination bands (see 5.25~\mum AIB)\\
\textbf{5.75} & 1739& & & & mw & & combination bands (see 5.25~\mum AIB) \\
& &  5.642 & 1772& \checkmark & mw & blend, symm &  \\
& &  5.699 & 1755& \checkmark & mw & blend, symm &   \\
& &  5.755 & 1738& \checkmark & mw & blend, symm &  \\
5.878 & 1701& & & \checkmark & mw, w & blend (5.75), symm &  combination bands (see 5.25~\mum AIB) \\[5pt]
\arrayrulecolor{lightgrey}\hline
 & \\[-5pt]
\multicolumn{8}{c}{\textbf{6.2~$\mu$m (1610~cm$^{-1}$) AIB} }\\[5pt]
6.024 & 1660& & &  & w, vw\tablefootmark{b} & blend (6.2), symm &  C=O stretch in conjugated carbonyl groups i.e., as quinones or attached to aromatic rings  \\ %
\textbf{6.2} & 1613 & & & & vs& asymm & pure aromatic CC stretching mode \\
& & $\sim$6.07 &$\sim$1647 & \checkmark & mw & blue shoulder\\
& &  6.212 & 1610&  & vs & asymm & \\
& &  $\sim$6.395 & $\sim$1564& \checkmark? & vw\tablefootmark{b} & blend   \\
& &  $\sim$6.50? & $\sim$1538?& \checkmark? & vw & slope change & \\[5pt]
\arrayrulecolor{lightgrey}\hline
 & \\[-5pt]
\multicolumn{8}{c}{\textbf{7.7~$\mu$m (1300~cm$^{-1}$) AIB complex} }\\[5pt]
6.638 & 1506 & & &  & vw & blend (6.711), symm & modes with a mixed character of CC stretching and CH in-plane bending \\
6.711 &  1490& & & \checkmark? & vw & blend (6.638), symm  & see 6.638~\mum AIB \\
6.850 &  1460& & &  & vw & blend (6.943), symm & CH deformation mode of aliphatic groups\\
6.943 &  1440& & &  & vw & blend (6.850), symm & CH deformation mode of aliphatic groups\\
\\
\textbf{7.7} & 1299  & & & & vs & &see 6.638~\mum AIB \\
& &   $\sim$7.05? &$\sim$1418? & \checkmark & vw\tablefootmark{b} & slope change& \\
& &   $\sim$7.10 &$\sim$1408 & \checkmark & vw\tablefootmark{b} & slope change &   \\
& &   $\sim$7.24 & $\sim$1381&  & vw\tablefootmark{b} & blend &   \\
& &   $\sim$7.43 & $\sim$1346&  & vw\tablefootmark{b} & slope change &   \\
& &   $\sim$7.626 & $\sim$1311&  & vs & blend &   \\
& &   \textbf{7.8} & 1290\\
& &   \hspace{.2cm}\rotatebox[origin=c]{180}{$\Lsh$}\,$\sim$7.743 & \hspace{.2cm}\rotatebox[origin=c]{180}{$\Lsh$}\,$\sim$1291& \checkmark & ms\tablefootmark{b} & blend    \\
& &   \hspace{.2cm}\rotatebox[origin=c]{180}{$\Lsh$}\,$\sim$7.775 & \hspace{.2cm}\rotatebox[origin=c]{180}{$\Lsh$}\,$\sim$1286& \checkmark & ms\tablefootmark{b} & blend\\
& &   $\sim$7.85 & $\sim$1274& \checkmark & ms\tablefootmark{b} & blend &   \\
 8.223 & 1216 & & & \checkmark?& vw\tablefootmark{b} & blend  & see 6.635~\mum AIB \\
 8.330 & 1200 & & & & vw, w\tablefootmark{b} & blend, symm & see 6.635~\mum AIB \\[5pt]
\arrayrulecolor{lightgrey}\hline
 & \\[-5pt]
\multicolumn{8}{c}{\textbf{8.6~$\mu$m (1160~cm$^{-1}$) AIB complex}}\\[5pt]
\textbf{8.6} & 1163 & &  & &s & symm & CH in-plane bending with large  CC stretching admixture  \\
& &  $\sim$8.46 & $\sim$1182 & \checkmark&  vw \tablefootmark{b}& slope change  &  \\
& &  $\sim$8.54 & $\sim$1171 & \checkmark&  vw\tablefootmark{b} & slope change &  \\
& &  8.60 & 1163 & &  s & symm &  \\
& &  $\sim$8.74? & $\sim$1144? & \checkmark&  vw\tablefootmark{b} & slope change \\
& &  $\sim$8.89? & $\sim$1125? & \checkmark&  w, vw\tablefootmark{b} & slope change &  \\[5pt]
\arrayrulecolor{lightgrey}\hline
 & \\[-5pt]
\multicolumn{8}{c}{\textbf{10--20~$\mu$m (500--1000~cm$^{-1}$) range\tablefootmark{c}}}\\[5pt]
 $\sim$10.95 & $\sim$913 & &  & & w & blue shoulder & solo CH out-of-plane bending   \\
 11.005 & 909 & & & &  mw, w\tablefootmark{b} & blend (11.2), symm & solo CH out-of-plane bending  \\
 \textbf{11.2} & 893& &  & & vs & asymm & solo CH out-of-plane bending \\ 
 & &   $\sim$11.207 & $\sim$892  & & vs& blend & \\
 & &  $\sim$11.25 & $\sim$889 & \checkmark& ms\tablefootmark{b}&  blend&   \\
 & &   $\sim$11.275? & $\sim$887?& \checkmark& mw\tablefootmark{b}& blend & \\
 \textbf{12.0}\tablefootmark{d} & 833\tablefootmark{d}~&  & & & ms, mw& blend (11.2) & duo CH out-of-plane bending \\
 & &  11.955 & 836 & & mw, w\tablefootmark{b}& &   \\
 & &  12.125?\,\tablefootmark{d} & 825?\,\tablefootmark{d}& \checkmark&  w\tablefootmark{b}& slope change &  \\
\textbf{12.7}\tablefootmark{d} & 787\tablefootmark{d}&  & & & s, ms & asymm & duo, trio CH out-of-plane bending\\
 & &   $\sim$12.38  & $\sim$808 & \checkmark& mw\tablefootmark{b}& blend&  \\
& &   $\sim$12.52? & $\sim$799? & & mw\tablefootmark{b}& blend &   \\
& &   $\sim$12.625 & $\sim$792 & \checkmark& mw\tablefootmark{b}& blend &   \\
& &   12.729 & 786 & \checkmark& s, ms& blend &   \\
& &  12.779 & 783 &\checkmark & s, ms& blend &   \\
& &   $\sim$12.98 & $\sim$770  & \checkmark& w\tablefootmark{b} & red shoulder&   \\
\textbf{13.5}\tablefootmark{d}& 741\tablefootmark{d}& & && mw, w & & quartet CH out-of-plane bending \\
& &   $\sim$13.50? & $\sim$738? & \checkmark& w, vw\tablefootmark{b}& blend \\
& &   13.55 &  & & mw,w& peak\\
& &   $\sim$13.62 & $\sim$734 & \checkmark& vw\tablefootmark{b}& blend\\
13.95?\tablefootmark{d}& 717?\tablefootmark{d}& & & & w, vw\tablefootmark{b}& & quintet CH out-of-plane bending, CCC skeletal \\
14.21\tablefootmark{d}& 704\tablefootmark{d}& & & & mw, w\tablefootmark{b}& & quintet CH out-of-plane bending, CCC skeletal \\
16.43 & 609 & & & & ms & asymm & CCC skeletal\\ 
\end{longtable}
\tablefoot{Columns: (1) Wavelength of AIB; nominal AIB names are listed in boldface; (2) Wavenumber of AIB; (3) Wavelength of component; (4) Wavenumber of component; (1)-(4) Tentative detection indicated by `?'; (5) New detection; (6) Peak intensity relative to the 3.3, 7.7 or 11.2 peak intensity for bands in the 3--4, 5--10, and 10--15 \mum range, respectively. The categories are vs ($>65$\%), s  ($>35$\%), ms ($>15$\%), mw ($>5$\%), w ($>2$\%), and vw ($<2$\%), where v=very, m=moderate, s=strong, and w=weak. The relative peak intensity of AIBs strongly depend on the employed continuum and spectral decomposition; (7) symmetric profile (symm); asymmetric profile (asymm);  blended with another band/component at x~\mum (blend (x)); change of slope (slope change); (8) vibrational assignment. \hspace{.5cm}
\tablefoottext{a}{From \citet{Peeters:nirspec}.}
\tablefoottext{b}{Estimated assuming the component is superimposed on other AIB component(s) such as, e.g., a blue or red wing or shoulder, i.e. we assume a local continuum. }
\tablefoottext{c}{We only list the strongest band in the 15--20~\mum region as it suffers from calibration issues (Appendix~\ref{sec:artefacts}). }
\tablefoottext{d}{Influenced by artefact(s) (Appendix~\ref{sec:artefacts}).}    }
\twocolumn

\section{The impact of residual artefacts in the data}
\label{sec:artefacts}

At the time of writing, there are two types of artifacts in the data that 
sometimes have appearances similar to AIBs.
Firstly, the constructive and destructive interference of layers in the MIRI detector arrays leads to periodic variations in the observed signal versus wavelength called ``fringes'' \citep{argyriou2020}. The latest versions of the \jwst\ pipeline and calibration reference files greatly reduce, but do not completely remove all fringes. These residual fringes can sometimes be difficult to distinguish from AIB components. 

Secondly, at the moment, the absolute flux calibration of \jwst\ \citep{gordon2022} is based on a single A star (private communication with the \jwst\ Help Desk). 
Intrinsic spectral differences between the star that was used for calibration and the
true spectrum of a given source can lead to artifacts in the observed spectrum of that source. Some artifacts are more complicated couplings between instrumental effects and incomplete absolute flux calibration.

Non-exhaustive list of wavelengths of artifacts:
\begin{enumerate}
\item \textbf{Excess emission:} 14.3, 14.35 $\mu$m
\item \textbf{Excess absorption:} $\sim$12.2, 13.71--13.77, 13.77--13.8, 13.85--13.95 $\mu$m
\item \textbf{Fringes:} 10--12 $\mu$m with an amplitude of about 5\% (affects red tail of 11.2)
\end{enumerate}

While future versions of the pipeline and reference files will fix fringes and calibration issues, we have developed a temporary fix for our analysis. We use MIRI-MRS observations of 10 Lac, which is an O9V star from the \jwst\ CALSPEC\footnote{\url{https://www.stsci.edu/hst/instrumentation/reference-data-for-calibration-and-tools/astronomical-catalogs/calspec}} program \citep{bohlin2014}. These data were reduced with the same versions of the \jwst\ pipeline and CRDS files that were used when reducing the Orion Bar data. Then we extracted the MIRI spectrum of 10 Lac using apertures whose radii increase linearly with wavelength. A comparison of the extracted spectrum and the CALSPEC model spectrum of 10 Lac revealed offsets in addition to fringes and absorption/emission artifacts. To deal with these offsets which vary from subband to subband, we performed linear regression on the extracted flux versus the model flux within a given subband, and then corrected the observed spectrum using the best-fit parameters for that subband. We then compute the ratio of the corrected, observed 10~Lac flux, to the model flux
\begin{equation}
C_\nu \equiv \frac{F_\nu^{\mathrm{10~Lac, ~obs.}}}{F_\nu^{\mathrm{10~Lac, ~model}}},
\end{equation}
where $F_\nu^{\mathrm{10~Lac, ~obs.}}$ is the (corrected) observed 10~Lac spectrum, and $F_\nu^{\mathrm{10~Lac, ~model}}$ is the model spectrum. Artifacts in the observed 10~Lac spectrum show up as deviations in $C_\nu$ away from 1.0.

We use $C_\nu$ to correct the Orion template spectra $F_\nu^{\mathrm{obs.}}$,
\begin{equation}\label{eq:corr}
    F_\nu^{\mathrm{corr.}}~\mathrm{[MJy~sr^{-1}]} = F_\nu^{\mathrm{obs.}} / \left( A_\mathrm{chan.,~band} (C_\nu - 1) + 1  \right),
\end{equation}
where $A_\mathrm{chan.,~band}$ is a parameter to adjust the amplitude of the correction for a given channel and subband. We apply corrections to Channel 2 LONG,  Channel 3 MEDIUM, and Channel 3 LONG, using $A_\mathrm{chan.,~band} = 0.5, 0.25, 1.0$, respectively (these factors were chosen manually). 
For all other wavelength ranges, no correction was applied either because no fringes were visible, or the correction introduced noise, making it difficult to tell if the fringes were actually corrected or simply buried in noise.

\begin{figure}
    \centering
    \resizebox{\hsize}{!}{\includegraphics{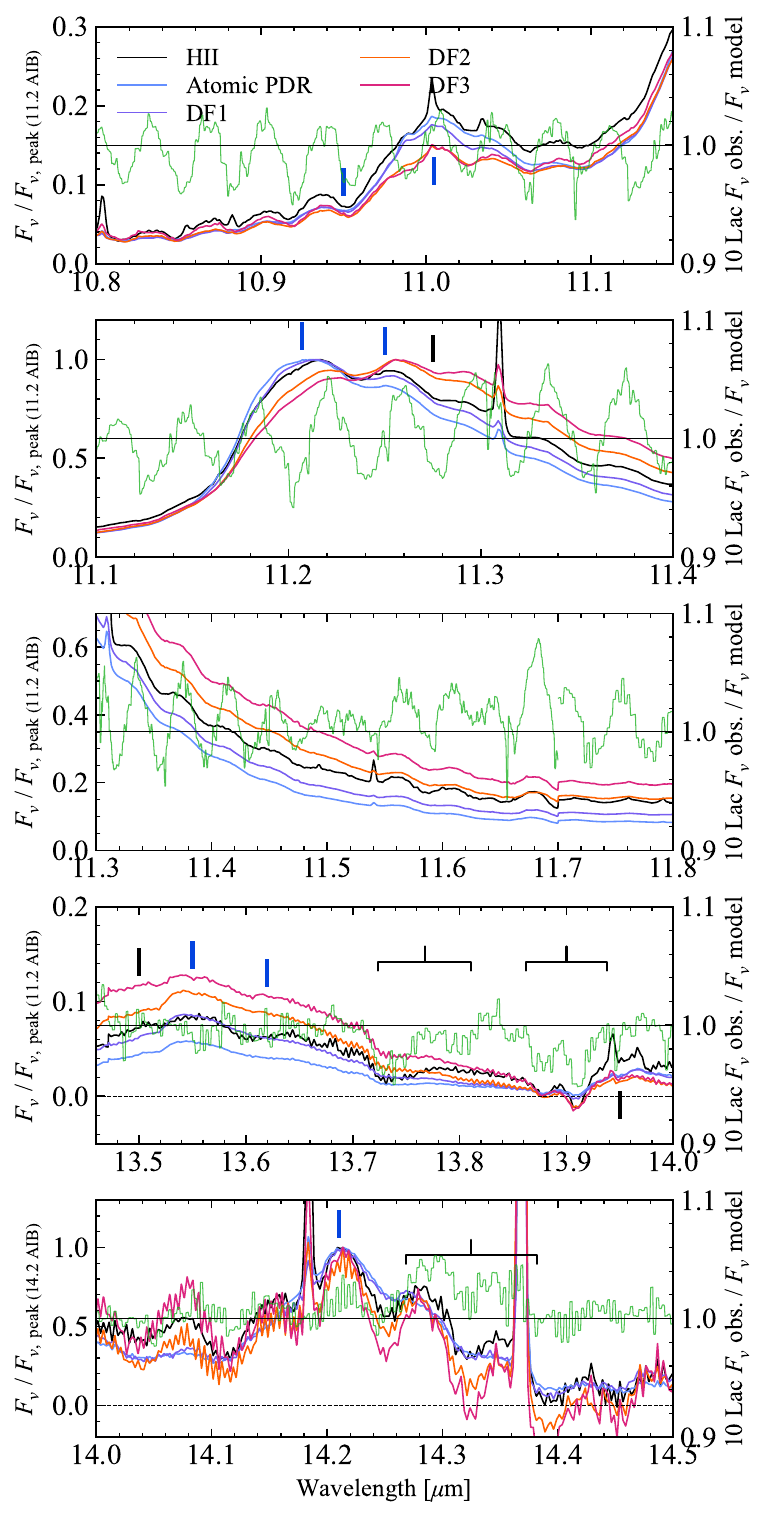}}
     \caption{Zoom-ins on continuum-subtracted template spectra (not corrected for calibration issues, i.e. $F_\nu^{\mathrm{obs.}}$ in Eq.~\ref{eq:corr}), compared with the MIRI-MRS spectrum of the absolute calibration star 10~Lac  divided by the model for this star (green). The positions of identified AIB features are shown with vertical lines. The bracketed regions indicate excess absorption or emission artifacts due to incomplete flux calibration (Appendix~\ref{sec:artefacts}). 
     }
    \label{fig:artefacts}
\end{figure}

\begin{figure}
    \centering
    \resizebox{\hsize}{!}{\includegraphics{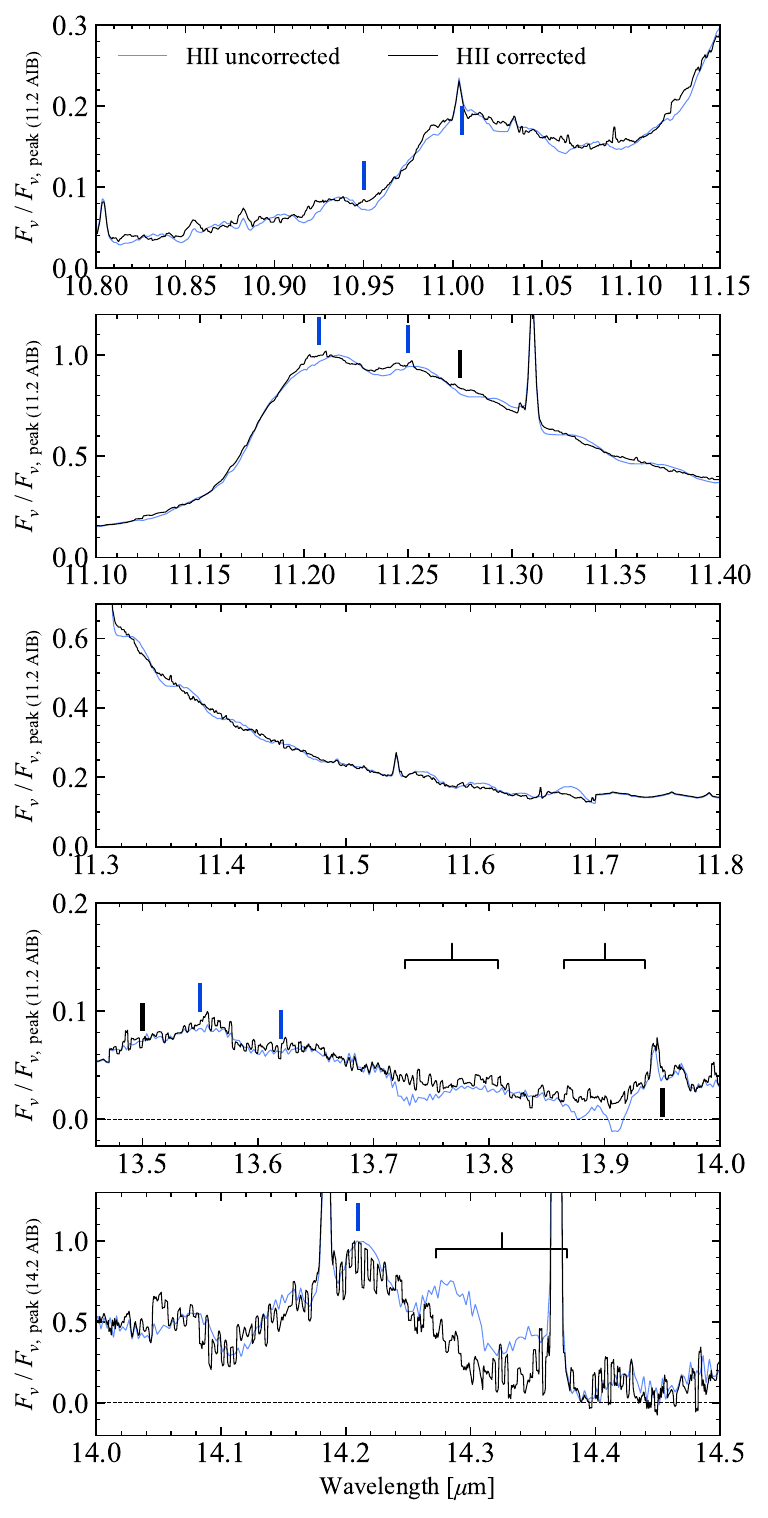}}
    \caption{$F_\nu^{\mathrm{corr.}}$ (black) and $F_\nu^{\mathrm{obs.}}$ (blue) for the \HII\ region template spectrum, showing improvements from the correction (Eq.~\ref{eq:corr}).
    The positions of identified AIB features are shown with vertical lines. The bracketed regions indicate excess absorption or emission artifacts due to incomplete flux calibration (Appendix~\ref{sec:artefacts}). 
    }
    \label{fig:corr_before_after}
\end{figure}

Examples of uncorrected template spectra are shown in Fig.~\ref{fig:artefacts}. 
Fringes and absorption/emission artifacts in 10 Lac (the green curve) 
are detected at the same wavelengths in the Orion data.  In Fig.~\ref{fig:corr_before_after} we show the \HII\ region template spectrum before and after applying the correction described above. The correction efficiently removes fringes (the first three panels),  excess absorption (fourth panel), and excess emission artifacts (bottom panel) which would have otherwise been incorrectly classified as AIB emission. We also conclude that the terraces in the 12.7~\mum AIB are real. We note however that some residual calibration issues remain, in particular at the longer wavelengths. We therefore restricted our analysis to wavelengths shorter than 15~\mum with the exception of the addition of the moderately strong 16.4~\mum AIB. We note that an artefact is also present at 18.9~\mum. This wavelength corresponds to the strongest band of C$_{60}$ emission \citep{Cami:10, Sellgren:10}. Unfortunately, due to the artefact, we are currently unable to confirm (or refute) the detection of C$_{60}$ in these templates.  In addition, the 12.2~\mum absorption feature is due to a spectral leak (estimated to be $\sim$3\% of the 6.1~\mum signal; \jwst\ Help Desk). This leak also affected the calibration star resulting in an incorrect photometric correction in the current pipeline. Due to the different spectral shape of Orion with respect to the calibration star, this resulted in the observed absorption feature and our applied correction is unable to correct this. This artefact influences the 12.0 and 12.7~\mum AIBs.

\section{Continuum determination}
\label{sec:cont_disc}
\begin{figure}
    \centering
   \resizebox{\hsize}{!}{\includegraphics{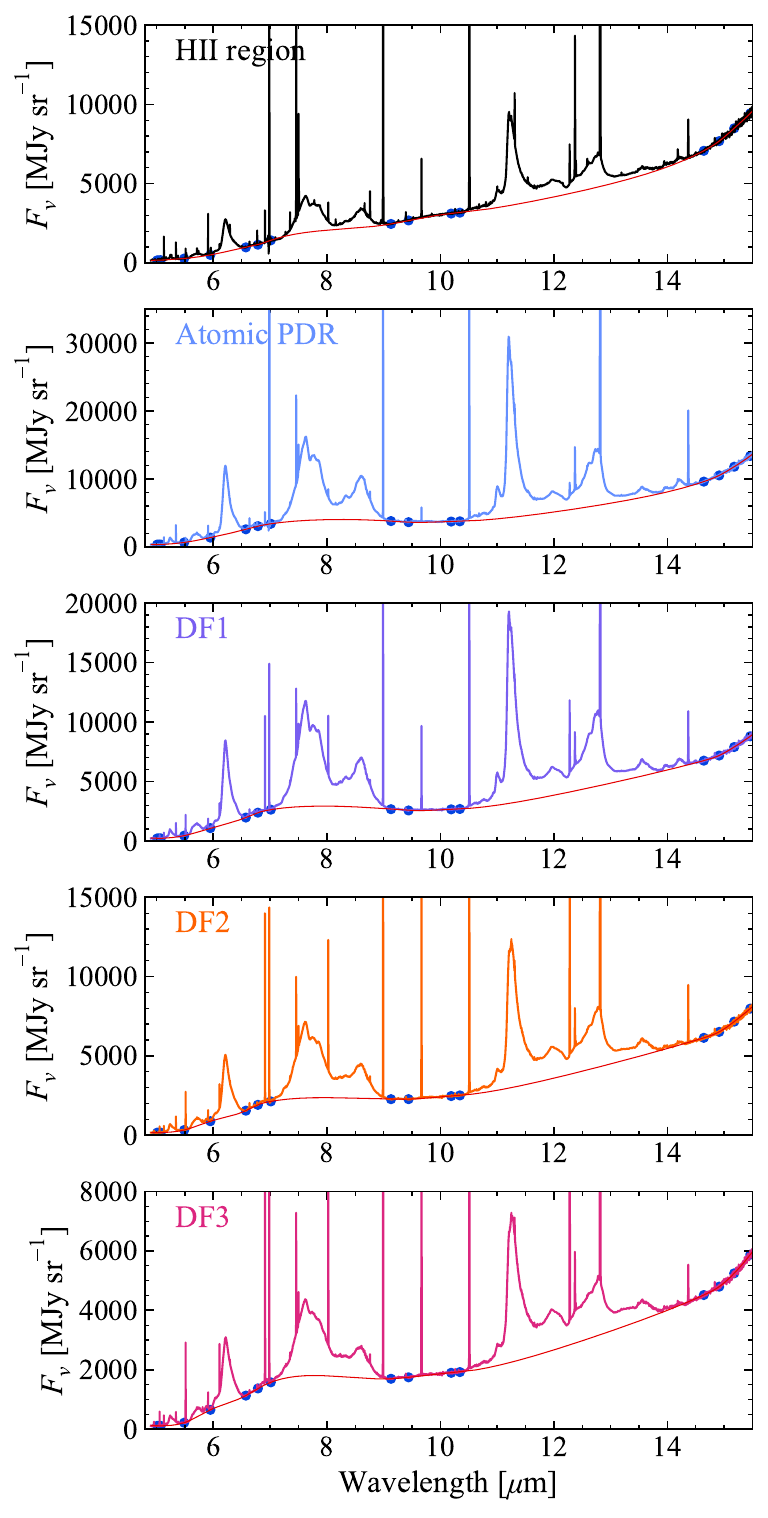}}
     \caption{The MIRI-MRS template spectra with their respective continua. The anchor points for each continuum are represented by filled circles. See Sect.~\ref{subsec:continuum} and Appendix~\ref{sec:cont_disc} for details.}
    \label{fig:cont}
\end{figure}

\begin{figure}
    \centering
   \resizebox{\hsize}{!}{\includegraphics{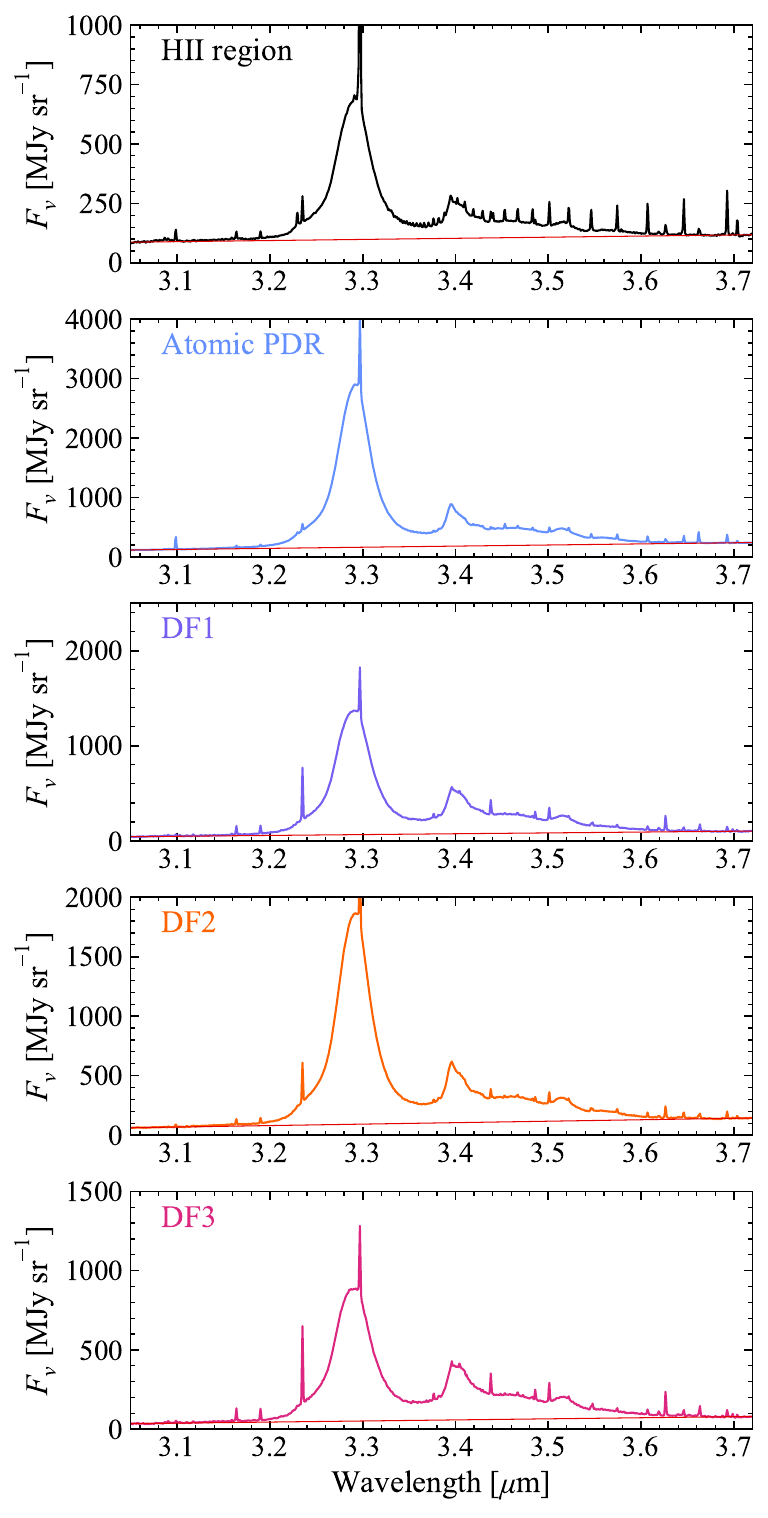}}
     \caption{The NIRSpec template spectra (grating g395h, filter f290lp) from \citet{Peeters:nirspec} along with their respective continua, zoomed in on the wavelength range that is relevant to the 3.3 $\mu$m AIB. See Sect.~\ref{subsec:continuum} and Appendix~\ref{sec:cont_disc} for details.
     }
    \label{fig:cont_nirspec}
\end{figure}

Here we show the continuum curves and the anchor points that were used to estimate them. Figure~\ref{fig:cont} shows the MIRI MRS template spectra and continua, while Figure~\ref{fig:cont_nirspec} shows the same but for NIRSpec data.

For the 3~\mum region (NIRSpec templates), we fit a linear continuum to the wavelength ranges [3.05, 3.07],
[3.667, 3.689], and [3.7068, 3.720], chosen to avoid emission lines. For the MIR (MIRI templates), we compute a cubic spline interpolation (tension of 1.0, IDL) anchored at selected wavelengths to estimate the continuum emission.  We selected anchor points that avoid all (weak) AIBs in all five templates. This resulted in anchor points at 5.011, 5.042, 5.073, 5.487, 5.948, 6.576, 6.785, 7.013, 9.131, 9.440, 10.191, 10.344, 14.642, 14.913, 15.178, 15.461, 19.349, and 19.632~\mum. The resulting continuum is shown in Fig.~\ref{fig:cont}. Our choice to use the same anchor points for all templates leads to slightly-overestimated continuum at 5.5~\mum in the DF~3 and at 7.1~\mum in the \HII\ region. Subtracting slightly-overestimated continuum from the templates will affect the resulting AIB emission in some cases.

\end{appendix}

\end{document}